\documentclass[preprint,12pt,authoryear]{elsarticle}

\usepackage{amssymb}
\usepackage{diffcoeff}  

\usepackage{lineno}

\usepackage{hyperref}

\usepackage[T1]{fontenc}

\journal{Astroparticle Physics}

\usepackage{graphicx} 
\usepackage[rightcaption]{sidecap}
\graphicspath{{Images/}}
\DeclareGraphicsExtensions{.pdf,.png}

\newcommand{\Figref}[1]{\mbox{{Figure \ref{#1}}}}%\mbox{}
\newcommand{\figref}[1]{\mbox{{Figure \ref{#1}}}}

\newcommand{\Secref}[1]{\mbox{{Section \ref{#1}}}}
\newcommand{\secref}[1]{\mbox{{Section \ref{#1}}}}

\usepackage{natbib}
\let\oldthebibliography\thebibliography
\renewcommand{\thebibliography}[1]{%
  \oldthebibliography{#1}
  \let\oldbibitem\bibitem
  \let\oldtextsc\textsc
  \def\oldbbland{et}
  \newcounter{authorcount}
  \def\bibitem[##1]##2{%
    \let\textsc\oldtextsc
    \let\bbland\oldbbland
    \oldbibitem[##1]{##2}%
    \let\textsc\mytextsc%
    \let\bbland\mybbland
    \setcounter{authorcount}{0}
  }
  \def\mybbland{\setcounter{authorcount}{0}\oldbbland}
  \def\dropetal##1.{ \bbletal}
  \def\mytextsc##1{%
    \oldtextsc{##1}%
    \stepcounter{authorcount}%
    \ifnum\value{authorcount}=5\relax%
      \expandafter\dropetal%
    \fi%
  }%
}
\begin{document}

\begin{frontmatter}


\title{A novel energy reconstruction method for the MAGIC stereoscopic observation}
\author[1,2]{Kazuma Ishio}\corref{cor1}
% \ead{ishio.kazuma@fis.uni.lodz.pl}
\ead{kazuma.ishio@umk.pl}
% \address[1]{Department of Astrophysics, The University of {\L}{\'o}d{\'z}, ul. Pomorska 149/153, {\L}{\'o}d{\'z} 90-236, Poland}
\address[1]{Institute of Astronomy, Faculty of Physics, Astronomy and Informatics, Nicolaus Copernicus  University, Grudzi\k{a}dzka 5,87-100 Toru{\'n}, Poland}
% \affiliation[1]{organization={aaa},
%             city={Lodz},
%             country={Poland}}

\author[2]{David Paneque}\corref{cor1}
\address[2]{Max-Planck-Institut f{\"u}r Physik, D-80805 M{\"u}nchen, Germany}
% \affiliation[label2]{organization={Max-Planck-Institute fuer Physik},
%             city={Munich},
%             country={Germany}}

\begin{abstract}

We report the successful development of a novel methodology of energy reconstruction for very high energy gamma rays detected with Imaging Atmospheric Cherenkov Telescopes (IACTs). This methodology, based on the machine learning algorithm Random Forest, and named \mbox{RF-Erec}, has been adjusted for being used with data from the Major Atmospheric Gamma-ray Imaging Cherenkov (MAGIC) stereo telescope system, which is a worldwide leading instrument for gamma-ray astronomy in the energy range from about 20\,GeV to beyond 100\,TeV.

The RF-Erec has been evaluated using different realistic scenarios with Monte Carlo simulated data and real observations from the Crab Nebula (the standard candle for the VHE gamma-ray community). This new methodology has been validated by the MAGIC software board, and it is implemented and ready-to-use in the MAGIC Analysis and Reconstruction Software (MARS).
 This new methodology, validated by the MAGIC software board, has been implemented and is ready for use in the MAGIC Analysis and Reconstruction Software (MARS). We demonstrate that, in comparison to the previous energy reconstruction methodology for MAGIC data, which relied on Look-Up-Tables (LUTs- Erec) and has been utilized in over 100 scientific publications over the last decade, RF-Erec significantly enhances the energy reconstruction of gamma rays. This improvement extends the capabilities of the MAGIC telescopes.  

Specifically, when quantifying the energy resolution with the width of a Gaussian fitted to the error distribution ({\it resolution-$\sigma$}),  the \mbox{RF-Erec} energy resolution-$\sigma$ is 20\% at 100 GeV and 11\% above 1 TeV for Zenith distances (Zd) below 35 degrees, while it is 20\% at 1\,TeV and 13\% above 10\,TeV for Zd above 55 degrees.  For a wide range of the observable energies, the improvement of energy resolution-$\sigma$, compared to LUTs-Erec, reaches roughly a factor of two, and the improvement is even larger for high Zd observations. Differently to many other works in the literature, our evaluation also xfconsiders the energy dispersion and the actual energy migration of events, where \mbox{RF-Erec} improves the performance of \mbox{LUTs-Erec} by factors of a few. The manuscript also demonstrates the importance of energy reconstruction methods with a small energy migration in order to prevent the appearance of artificial spectral features.  These artifacts are particularly important at the high end of the gamma-ray spectra, where a few extra high-energy photons could have dramatic consequences for studies related to the EBL attenuation, Lorentz invariance violation, or searches for Axion-like-particles.

\end{abstract}

\begin{keyword}

Very High Energy gamma ray \sep 
Cherenkov Telescopes \sep 
Energy reconstruction \sep 
Random Forest




\end{keyword}
\end{frontmatter}
% \linenumbers
% \begin{linenumbers}
%   \shortcites{Aleksic2016,Aliu2008}
\section*{Introduction}
%----  (Rough)
% \subsubsection{MAGIC telescopes and detection}
MAGIC is a system of two 17 m-diameter Imaging Atmospheric Cherenkov Telescopes (IACTs) located at the Roque de los Muchachos Observatory on the Canary Island of La Palma, Spain, at an altitude of 2200 m a.s.l.\citep{Aleksic2016, MAGIC2016upgrade2}. The MAGIC telescopes perform gamma-ray astronomy in the energy range that goes from about 20 GeV to beyond 100 TeV. 
%main target of the system is to observe very-high-energy (VHE, energy $\gtrsim$ 100 GeV) gamma rays. %----   (Technical detail)
The gamma-ray detection method of IACTs is to record the Cherenkov radiation from an extensive air shower (EAS), which is a cascade of secondary particles initiated by a high energy particle such as a gamma-ray photon, when it enters the atmosphere. By analysing the images, the properties of primary particles, namely the particle species, incoming direction and energy, can be reconstructed. 
%The MAGIC was designed especially for the lowest energy possible for IACTs by optimizing various specifications. %; e.g. the large reflector area, the sensitive and high-timing-resolution camera, and the distance between the telescopes. 
% The achieved energy threshold is as low as $\sim$50 GeV at the trigger level for zenith angular distance (Zd) of observation direction at below 30 deg in dark conditions (without moon) \citep{MAGIC2016upgrade2}. Using the so-called sum-trigger, it is possible to reach an even lower energy threshold \citep{Dazzi2021}. 
In the case of observations performed within 30 deg of zenith angular distance (Zd) during moonless nights, the MAGIC telescopes can measure gamma-ray energies down to 50\,GeV for sources like the Crab Nebula \citep[e.g., see][]{MAGICstereoPreUpgrade}.  Additionally, when using the Sum-Trigger-II system (instead of the standard trigger system), the MAGIC telescopes increase the sensitivity for low-energy showers \citep{Dazzi2021}, and is able to produce scientific results with energies as low as 15\,GeV \citep{Acciari2020h}

% \citep[e.g., see][]{2015JHEAp...5...30A}. => 
% Measurement of the Crab Nebula spectrum over three decades in energy with the MAGIC telescopes

%Aliu2008, 
% Another notable submodule is the drive system with a fast repositioning speed in order to react to alerts from other detectors of rapid transient events such as Gamma-Ray Bursts. 

%\subsubsection{The energy estimator history in MAGIC}
In early times of MAGIC when there was only one telescope, the algorithm of the energy estimator was based on Random Forest (RF) \citep{Albert2008_hadronness}. With the introduction of the second telescope, the geometric stereo reconstruction of the shower became possible \citep{MAGICPerformancePaper2012}. In order to use this information for the gamma-ray energy estimation, the Look Up Tables method (LUTs-Erec), based on a simplified shower model, was introduced and it has been used over the last decade in the MAGIC standard analysis \citep{MAGIC2016upgrade2}. In this paper we present a new gamma-ray energy reconstruction method based on Random Forest that can be used with stereo observations, and that substantially improves the peformance of the MAGIC telescopes.  

%showed the best performance. It has been used over the last decade in the MAGIC standard analysis \citep{MAGIC2016upgrade2}. 
%When the telescopes system was upgraded by introducing the second telescope, the system started stereo observations \citep{MAGICPerformancePaper2012}. This improved significantly the estimation of the distance to the shower maximum, thus the correspondent upgrade was needed also for the data analysis including the energy estimator. At that time, the LUT method showed the best performance. It has been used over the last decade in the MAGIC standard analysis \citep{MAGIC2016upgrade2}.

 The paper is organized as follows: \secref{sec:EnergyEstimator} describes the energy reconstruction methods that have been used in MAGIC, together with a brief overview of those used in the other major IACTs. \secref{sec:NewEstimator} introduces the new method for energy reconstruction, based on Randon Forest, and \secref{sec:Verification} details the validation of the method, using three different approaches. In \secref{sec:Performance}, the performance of the new method is reported, together with a detailed comparison with the previous energy reconstruction method. At last, in \secref{sec:Conclusions}, we provide some concluding remarks and point out fundamental problems in deriving certain physics results when using an energy reconstruction with a poor resolution, or large spill-over fraction.

\section{Energy estimation methods}\label{sec:EnergyEstimator}

In the gamma-ray observation with IACTs, the atmosphere acts as an inhomogeneous calorimeter; the energy of the gamma ray is deposited in the atmosphere and a fraction of it is radiated in the form of Cherenkov light, that can be detected and inferred as the brightness of the EAS. 

Therefore the estimation of the energy of the incoming gamma ray is based on the brightness of the image (recorded in the camera of the IACT) corrected for instrumental (impact distance, telescope efficiency, truncation in the camera field of view) and environmental (zenith angle, atmosphere) effects. The reconstruction of the incoming gamma-ray event is naturally improved when the observation of the EAS is performed by multiple IACTs, the so-called stereo observation. In the following we describe the details of MAGIC energy reconstruction. We begin with the need of training samples, then we continue with the MAGIC official gamma-ray reconstruction process featuring the official energy reconstruction before the application of the method presented here, LUTs-Erec. At last, we also introduce other energy estimators.

% \subsubsection{MC}
%---- (MC) 
\subsection{Training aspects}

A number of simulated events, called train samples, is needed in order to generate a gamma ray energy estimator. It provides the relation between a gamma ray and its measured property with the telescopes to a supervised learning of an energy estimator. 
MAGIC utilizes the standard MAGIC simulation package \citep{Majumdar2005a, Carmona2007}, with the gamma-ray showers generated using the CORSIKA code \citep{Heck1998}. In this study, we use the version 6.990. Monte Carlo (MC) simulated gamma rays are generated in a sufficiently smooth distribution of energy, pointing direction, incoming direction and arrival location on the ground. The simulation ranges and the statistics are adjusted depending on the target ranges, and the productions are shared among the collaboration. 

The most standard setting is for the case of a point-like source observed at Low Zd (5 deg to 35 deg) with a wobbling angle \citep{Fomin1994a} of  0.4 deg. %where the pointing direction is offset from the incoming direction by 0.4 deg. 
This data set consists of roughly $3 \cdot 10^5$ triggered events from roughly $5 \cdot 10^6$ gamma ray events generated. To be generally used for any observation of this kind, the pointing directions are equally distributed in Az and given Zd range. The wobbling offset direction of the source in the field of view (FoV) is equally distributed as well in the ring-like region, to take into account the tracking and the change of wobbling position. The thickness of the ring is determined by the pointing accuracy,  which is of the order of 0.02 degrees , as discussed in section 5.2 of \citet{MAGIC2016upgrade2}.
The arrival locations on the ground are chosen in uniform distribution within a circular region with a radius of 350 m around the center of the telescopes system in the plane perpendicular to the shower axis.  The energy distribution spans from 10 GeV to 30 TeV \footnote{
  The energy ranges used are  10GeV-30TeV (for the Zd ranges 0-35deg and for 35-50 deg), 40GeV- 80 TeV (for the Zd range 55-65 deg) and 400GeV-160 TeV (for the Zd range 70-80 deg).
}, distributed with a power-law shape with a differential index of -1.6.  %of differential index -1.6.

\subsection{Image parameters}
% \subsubsection{parametrisation}
%---- (Parametrisation)
By detecting the photons from an EAS with the two telescopes, we obtain two images of the EAS seen from different locations. These images are not only the distributions of light intensity (deposited charge) over the pixels, but also that of  photon arrival times. 
The pixels dominated by noise are identiﬁed and excluded by image cleaning techniques \citep{MAGIC2016upgrade2}, and the signal in the pixels surviving the cleaning form an image that is quantified through an analytical parametrization.

Among the parameters analytically calculated from the images, the ones used for the  energy reconstruction are the following: 

\begin{description}
\item[Length, Width]
These are part of the well-known Hillas parameters \citep{Hillas1985}, modelling the image shape as an ellipse. Length and Width are calculated as the second moment of the charge distribution along the major and minor axis. 

\item[Size]
 Also as a part of the Hillas parameters,  Size is the total charge in the unit of number of photoelectrons (phe), summed over all pixels surviving the image cleaning. 

\item[Leakage (Leakage1, Leakage2)]
As an auxiliary parameter, Leakage characterises the truncation of the image, by quantifying the fraction of the charge deposited in the pixels on the edge of the camera's FoV with respect to Size, namely the total charge. There are two kinds of Leakage defined; Leakage1 is for the outmost row on the edge, and Leakage2 is for the second row. 

\item[TimeGradient]
As for the arrival time distribution, the most important parameter is TimeGradient, which is the gradient of the arrival time distribution with respect to the projected location of the pixels along the major axis \citep[see further details in][]{MAGICtimegrad}. The sign and value of the gradient depend, among others, on the impact distance.
\end{description}

The image parameters are firstly used for determining the incoming direction of the gamma ray, which is essential for calculating the shower geometry. The shower geometry, especially the distance to the EAS and to its shower axis is very helpful for the energy reconstruction, as well as the scoring of the events for discriminating the gamma ray events from other background events.

 \subsection{Incoming direction reconstruction and shower geometry parameters}
The incoming direction is estimated using the so-called Disp method (see fig.~5 in \citealp{MAGICPerformancePaper2012}). The method is based on the Disp parameters for individual camera FoVs, each of which is the angular distance from the Center of Gravity position (CoG) to the incoming direction of the primary gamma ray. Disp is estimated from the image parameters. As an independent simple method, the incoming direction is calculated also by the classic method as the intersection of the shower axes \citep{Daum1997}. 
Based on the above mentioned incoming directions, the shower geometries are calculated and parametrized. We discriminate these two kinds of geometries by adding the prefixes Disp- and Classic- to the shower geometry parameters respectively. The Disp- geometry performs considerably better than Classic- geometry. 

The most important shower geometry parameters are Impact (for each telescope) and MaxHeight. The former is the lateral distance from a telescope to the shower axis, and the latter is the height at which the Cherenkov emission from the EAS is maximum. The parameter Disp is correlated to both Impact and MaxHeight. The distributions of the reconstructed parameters (and in particular MaxHeight and Impact) depend on the incoming direction of the gamma ray and, with the goal of constructing robust estimator, one needs to consider (or even remove/mitigate) this dependency. There are other shower geometry parameters used in the standard analysis for this purpose, by constructing from the MaxHeight and Zd combined with the location specific information such as atmosphere density and geomagnetic field. The parameters CherenkovDensity and CherenkovRadius indicate the longitudinal and lateral information respectively. They are to characterise the distance to the shower maximum in terms of the Cherenkov radiation, by a 86 MeV electron at the shower maximum. Below this energy, in the air, the electrons lose their energy quickly by ionization and no longer contribute to the shower. CherenkovRadius is the radius of illumination on the plane perpendicular to the pointing direction at the center of the two telescopes, and CherenkovDensity is the photon density of illumination on the plane. 

The reconstructed shower geometry can also integrate the effect of the geomagnetic field. CosBSangle is the cosine of angular distance of the incoming direction from the geomagnetic field direction, and is used to account the deflection of the secondary particles, which have a direct impact on the Cherenkov radiation \citep{MAGICgeomagnetic}.

\subsection{The previous energy estimator used in the MAGIC telescopes}\label{sec:PreviousEstimator}

The MAGIC collaboration has been estimating (during more than a decade) the energy of the gamma rays from the MAGIC stereo observations using a method based on Look-Up-Tables \citep{MAGIC2016upgrade2}. In this manuscript, we name this method LUTs-Erec. This method is based on two telescope-wise energy estimators, where each of them takes the image parameters for the correspondent telescope and the shower geometry parameters. And once the telescope-wise estimations are obtained, they are averaged weighted with uncertainties. 

The procedure of obtaining the telescope-wise estimation consists of two stages. In the first stage, the Size parameter is modified via empirical formulae using the following parameters:

\begin{itemize}
\item Image truncation correction \\
For the events with truncated images, Size needs to be compensated with the missing amount of light content outside the field of view. Size is multiplied by
%1-4*x*x
$$
\frac{1}{1 - 4\,x^2}
$$
where $x=$ Leakage2.

\item Zd correction\\
The larger the Zd, the longer distance to the shower maximum point, thus the dimmer image results in smaller Size. To compensate this, Size is multiplied by 
%0.97*pow(x,-0.3)/(1-pow(1-x,2.25))
$$\frac{0.97\,x^{-0.3}}{1-(1-x)^{2.25}}$$
where $x = \cos{Zd}$.

\item Magnetic correction \\
In the presence of the geomagnetic field, the acceleration acts in opposite directions on electrons and positrons in the electromagnetic cascades. The separation leads to a stretched distribution of light pool and to a drop of intensity.
Size is multiplied by
%0.93+0.2*sqrt(1.-pow(x,2))
$$0.93+0.2\sqrt{1-x^2}$$
where $x=$ CosBSangle.

\item Distance correction\\
Finally the Size is scaled taking into account the distance. To this end, CherenkovDensity is used as the chacterization of the density drop. 
Size is multiplied by
$$\frac{1}{x}$$
where $x=$ CherenkovDensity.
\end{itemize}

In the second stage, the conversion factor from the corrected Size to energy is retrieved by searching through a two-dimensional LUT, which has the corrected-Size as one dimension, and the Disp-Impact parameter as the other dimension.  The corrected-Size is expressed as $\sqrt{\log_{10} (\mbox{corrected-Size})}$ %square root of the logarithm
 in order to optimize the binning, while the Disp-Impact is expressed in the unit of Disp-CherenkovRadius so that it can take into account that the light pool radius changes according to the longitudinal distance. 

The estimate in each grid of LUT is determined by averaging the training samples filled in the correspondent grid. The shower geometry parameters used in the training have different origins. The CherenkovDensity and CherenkovRadius are calculated directly from the shower images.  On the other hand, the Impact is extracted directly from the simulation, and the CosBSangle is calculated from the true incoming direction as the direction of primary particle injection. For the application of the LUT energy reconstruction to the actual observational data, one needs to replace the MC information by estimates of these quantities.

\subsection{Other energy estimators used in IACTs}

%%%%%%%%%%%%%%%%%%%%%%%%%%%%%%%%%%%%%%%%%
%Holler2015 : Photon Reconstruction for H.E.S.S. Using a Semi-Analytical Shower Model
%https://arxiv.org/abs/1509.02896
Along with MAGIC, the other major ground-based observatories at present with a system of more than one imaging Cherenkov telescope %\citep{Funk2015a} 
are the High Energy Stereoscopic System \citep[H.E.S.S.,][]{TheHESSCollaboration2006} and the Very Energetic Radiation Imaging Telescope Array System \citep[VERITAS,][]{Meagher2015}. 
%Both of them have the same basis of detection that the multiple telescopes view the shower from different angles. %, further suppressing the background. 
In the energy reconstruction, they adopt different strategies. 

%%%%%%%%%%%%%%%%%%%%%%%%%%%%%%%%%%%%%%%%%
%  VERITAS
%%%%%%%%%%%%%%%%%%%%%%%%%%%%%%%%%%%%%%%%%
%Park2015 : https://arxiv.org/pdf/1508.07070.pdf
Standard analyses for VERITAS use look-up tables derived from simulated events to reconstruct the energy of incident air showers based on the strength of the signal and the distance of the air shower from the telescope \citep{Park2015}. The energy bias is close to zero above around 200 to 500 GeV depending on the cut condition, and below a few tens of TeV. The energy resolution %, defined as the 68 \% containment width around the median value of the energy bias distribution, 
is about 15 \% - 20 \%, with a worsening of the resolution for energies larger than 10 TeV \citep{Park2015}.  

%%%%%%%%%%%%%%%%%%%%%%%%%%%%%%%%%%%%%%%%%
%  HESS (ImPACT)
%%%%%%%%%%%%%%%%%%%%%%%%%%%%%%%%%%%%%%%%%
% Parsons2015: HESS II Data Analysis with ImPACT
% https://ui.adsabs.harvard.edu/abs/2015ICRC...34..826P/abstract 
% https://arxiv.org/pdf/1509.06322.pdf

%%%%%%%%%%%%%%%%%%%%%%%%%%%%%%%%%%%%%%%%%
% Parsons2014: A Monte Carlo template based analysis for air-Cherenkov arrays
% https://www.sciencedirect.com/science/article/pii/S0927650514000231?via%3Dihub
H.E.S.S. adopts two standard analyses, called Model++\citep{DeNaurois2009} and Image Pixel-wise fit for Atmospheric Cherenkov Telescopes  \citep[ImPACT,][]{Parsons2014, Parsons2015}. They are based on the likelihood fitting of camera pixel amplitudes to an expected image template, which are generated by a dedicated semi-analytical code, or directly generated from Monte Carlo simulations, for the Model++ and the ImPACT, respectively.

For the events with standard ImPACT method standard cuts, the energy bias is close to zero above 500 GeV, and the resolution is better than 10 \% above 1 TeV \citep{Parsons2014, Parsons2015}. 

Note that the definitions of energy bias and resolution used in these publications are somewhat different, in both the event selections and the computation. For instance, while they are both based on the distribution of  the  fractional deviation of the reconstructed energy from the simulated energy, \citep{Park2015} adopts the resolution as  a \mbox{68 \%} containment radius around the median value, and  \citep{Parsons2014} uses the RMS of the distribution.

%%%%%%%%%%%%%%%%%%%%%%%%%%%%%%%%%%%%%%%%%
% DL
%%%%%%%%%%%%%%%%%%%%%%%%%%%%%%%%%%%%%%%%%
%These days, deep convolutional neural networks (CNNs) are being explored as a promising method for IACT full-event reconstruction \citep{Miener2021}. 
These days, deep convolutional neural networks (CNNs) are being explored as a promising method for IACT full-event reconstruction. 
There have been some remarkable attempts \citep{Nieto2021, De2022, Jacquemont2021}, based on a technique to interpret a hexagonal pixel layout into cartesian coordinate \citep{Nieto2021a}. In the latest study done within the MAGIC collaboration \citep{Miener2021}, the pixel-wise image information is fed to the estimator after the image cleaning. The performance achieved with this CNN method, called CTLearn, is comparable to the one that we achieve with the method we describe in this paper, as depicted in the first panel of \figref{fig:GeneralizationErr}. However, a drawback of this methodology is the demand for expensive computational resources  at the training stage, which has to be repeated many times due to different MC datasets for different periods of time.  The production of an estimator is required not only for the observation settings like pointing directions, but also when there is a change in performance of the telescopes. Therefore an estimator that is computationally less demanding, and more flexible in its usage, may play an important role to enable  swift modifications in the analysis.We also note recent works that use graph neural networks (GNN) for signal/background separation on IACT data (e.g., \citep{Glombitza2023}) . This strategy does not require a mapping from hexagonal pixels into cartesian space and looks very promising.

\section{The new energy estimator, RF-Erec}\label{sec:NewEstimator}

\subsection{Introduction to Random Forest }
The new energy reconstruction method, RF-Erec, is based on the usage of Random Forest (RF)\citep{Breiman2001}. The Random Forest method is based on a collection of decision trees. In the typical decision tree method, one processes an event, down through a tree, for estimating its target value. The tree consists of a chain of binary decisions called nodes each of which evaluates whether an input variable is above or below a threshold. A decision tree is constructed with a set of train samples, by the iteration of splitting them on the best split position of the best input variable into two groups with more similar target values. As the splitting process repeats, the number of events in the training sample gets reduced. The repetition stops after a number of times, called tree depth, or when the number of events in the sample is smaller than a given number, called minimum node size. The events in the sample that remain after the last split, called last node, determine the estimation value from the population. This tree decision process is very sensitive to over-training, that happens when the minimum node size is very small and the final decision follows also the outliers in the training sample. The RF method was introduced to mitigate this over-training. The RF consists of a collection of decision trees, built up with two elements of randomization: the bootstrap aggregating (bagging), and the random selection of the input variables used to cut the sample. The first one, the bagging, is used to randomize the train sample. This is done through the generation of a set of randomly picked train samples with the same number of events as the original one, but with some events duplicated. The second one, the random choice of the input variables in each split, consists in selecting the best input variable from a limited number of the input variables, that are randomly chosen in each split. The final estimation of the target value is done by the ensemble of those from such randomly produced trees. Some of the individual trees may grow in a way that they are biased towards something (or affected by a few outliers); but the average of the estimates from all the trees is expected to be a robust (unbiased one). In this way, by construction, the RF is designed to be much less sensitive to outliers in the training sample, and hence more robust against over-training. 

Among a number of machine learning algorithms, RF is nowadays rather traditional. It faces greater challenges with high-dimensional data that includes a large number of noisy variables compared to state-of-the-art machine learning techniques such as convolutional neural networks. For example, when analyzing an image with RF, it requires selecting and deriving features for each object, making it more complex. However, if the features that relate to the target value are well quantified and used as input variables, RF can be an efficient method and it is one of the best methods \citep{Han2018,Olson2017}. In addition, RF is an appealing choice due to its low computational cost \citep{Han2018}, clear interpretability, and robustness, particularly its resilience to noise and outliers which is a typical challenge in a neural network based method \citep{Montavon2012}. More detailed and extensive discussion is seen in \citep{Hastie2009} for general cases for example, and in \citep{Ishio2020} for this study.

\subsection{Selection of the input variables}
We selected the input variables from the event parameters based on the insights on the EAS and the telescopes system, in order to benefit from the RF's ability as much as possible by avoiding noise variables as well as to mitigate unknown systematic effects \citep[details in][]{Ishio2020}. The selection followed the idea to construct the previous method, the LUTs-Erec, however extended to all the relevant parameters. We listed 21 parameters, from which 11 are parameters that had not been used in LUT-Erec. The list is shown in the Table 1. For taking into account the truncation of the image, both Leakage1 and Leakage2 are selected, while the LUTs-Erec uses only Leakage2. As for parameters indicating the distance to the shower, the LUTs-Erec used only Impact, CherenkovDensity, CherenkovRadius and Zd, while the RF-Erec used, in addition, Disp, MaxHeight, TimeGradient, Width and Length. In the RF-Erec, all the geometrical parameters use Disp- values.

\begin{table}[htbp]%[H]
\centering
\caption{The event parameters used for the energy estimators.  } \label{tbl:variables}
\begin{tabular}{l l| c | c }
\hline 
Event parameter    &&\mbox{RF-Erec} &  \mbox{LUTs-Erec}\\ 
\hline \hline 
Size    &(tel.1/2)  & •  & • \\ 
\hline 
Leakage1 &(tel.1/2) & • &   \\ 
\hline 
Leakage2 &(tel.1/2) & • & • \\ 
\hline 
Time Gradient &(tel.1/2) & • &   \\ 
\hline 
Width         &(tel.1/2) & •  &   \\ 
\hline 
Length        &(tel.1/2) & •  &   \\ 
\hline
\hline 
Disp          &(tel.1/2) & •  &   \\ 
\hline 
Disp-Impact   & (tel.1/2) &  •&   \\ 
True-Disp-Impact   & (tel.1/2) &   & • \\ 
\hline
Disp-MaxHeight   &  &  •&   \\ 
\hline
Disp-CosBSangle  &&  • &   \\ 
True-Disp-CosBSangle &&   &  • \\ 
\hline 
\multicolumn{2}{l|}{Disp-CherenkovDensity}
                  & • &   \\ 
\multicolumn{2}{l|}{Classic-Disp-CherenkovDensity}
                  &   & • \\ 
\hline 
\multicolumn{2}{l|}{Disp-CherenkovRadius}
                  & • &   \\ 
\multicolumn{2}{l|}{Classic-Disp-CherenkovRadius}
                  &   & • \\ 
\hline 
\multicolumn{2}{l|}{Zd of telescope pointing}
                  & • & • \\
\hline 
%\multicolumn{2}{l|}{Classic-Shower maximum height} & •  & •  & •  & • & {\color{red}•} & • \\ 
%\hline 
\end{tabular}
\begin{quote}
\footnotesize
The list of the event parameters used as input variables for the energy estimators, RF-Erec and LUTs-Erec. The geometrical parameters are indicated with the basis direction reconstruction methods as prefixes; Classic-, Disp- and True- (see the main text for the details). The ones with two prefixes indicate that the different reconstruction bases are used in the construction and the application of estimators, represented by the first and the second prefixes. 
\end{quote}
\end{table}

\subsection{Use of True- versus Reconstructed- parameters in training}
\label{sec:true-vs-recon}

As mentioned in \Secref{sec:PreviousEstimator}, the true MC values are used for some of the input variables in the training of LUTs-Erec.  The replacement of geometrical parameters by True- values has been especially common among the estimators for the IACTs. %also in other cases \citep{HESS2018}
 By learning the relation between the true geometry of the shower and the target value, the estimator was expected to avoid the smearing of the estimator response by these geometrical variable error. However, the difference of the parameters between simulation and measurement can appear and have negative effects when the bias in the measured parameter is larger than the statistical fluctuations. In our study with the application of the RF for the energy estimation in the MAGIC data, we found out that this strategy can introduce features in the gamma-ray spectra, such as wiggles  \citep[see Fig.~E.11 in ][]{Ishio2020}.  Because of that, in order to avoid these systematic effects, we decided to not follow this strategy in the development of RF-Erec. 

 \subsection{Event Selection Strategy}
The event selection for train samples was also updated from the LUTs-Erec. The LUTs-Erec needs an event selection before training because the limited range of validity of some of the parameters used for correcting the Size with a empirical formulae. Therefore, the optimization of the performance included an event selection that removes a non negligible fraction of the events.  This  event selection has two parts:

\begin{itemize}
\item The selection that is common in the two LUTs of the two telescopes (MAGIC-1 and MAGIC-2)\\
(All the parameters used are the geometrical parameters by Classic incoming direction reconstruction)
    \begin{itemize}
    \item Valid direction reconstruction
    \item Theta2 $<$\,0.1\,deg$^2$ \\
    (where Theta2 is the squared angular distance between the reconstructed and true incoming directions) 
    \item CherenkovRadius $>$ 40\,m
    \item CherenkovDensity $>$ 0 
    \end{itemize}
\item The  selection cuts applied to parameters  computed with single-telescope quantities
    \begin{itemize}
    \item Size between 25 and $2\cdot 10^5$\,phe
    \item TrueImpact normalized by CherenkovRadius $<$\,3.5 \\
     (where CherenkovRadius is from Classic incoming direction reconstruction)
    \item  Leakage1\,$<$\,0.2
    \end{itemize}
\end{itemize}

\begin{figure}[htbp]
\centering
\includegraphics[width=0.8\textwidth]{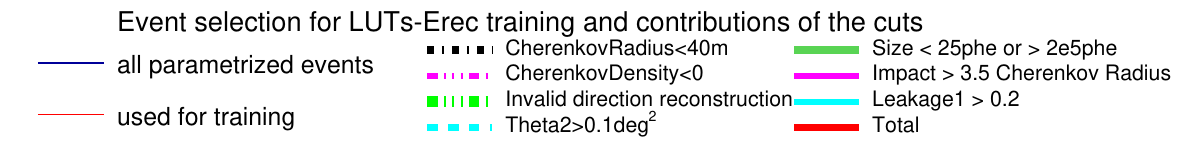}
\includegraphics[width=0.9\textwidth]{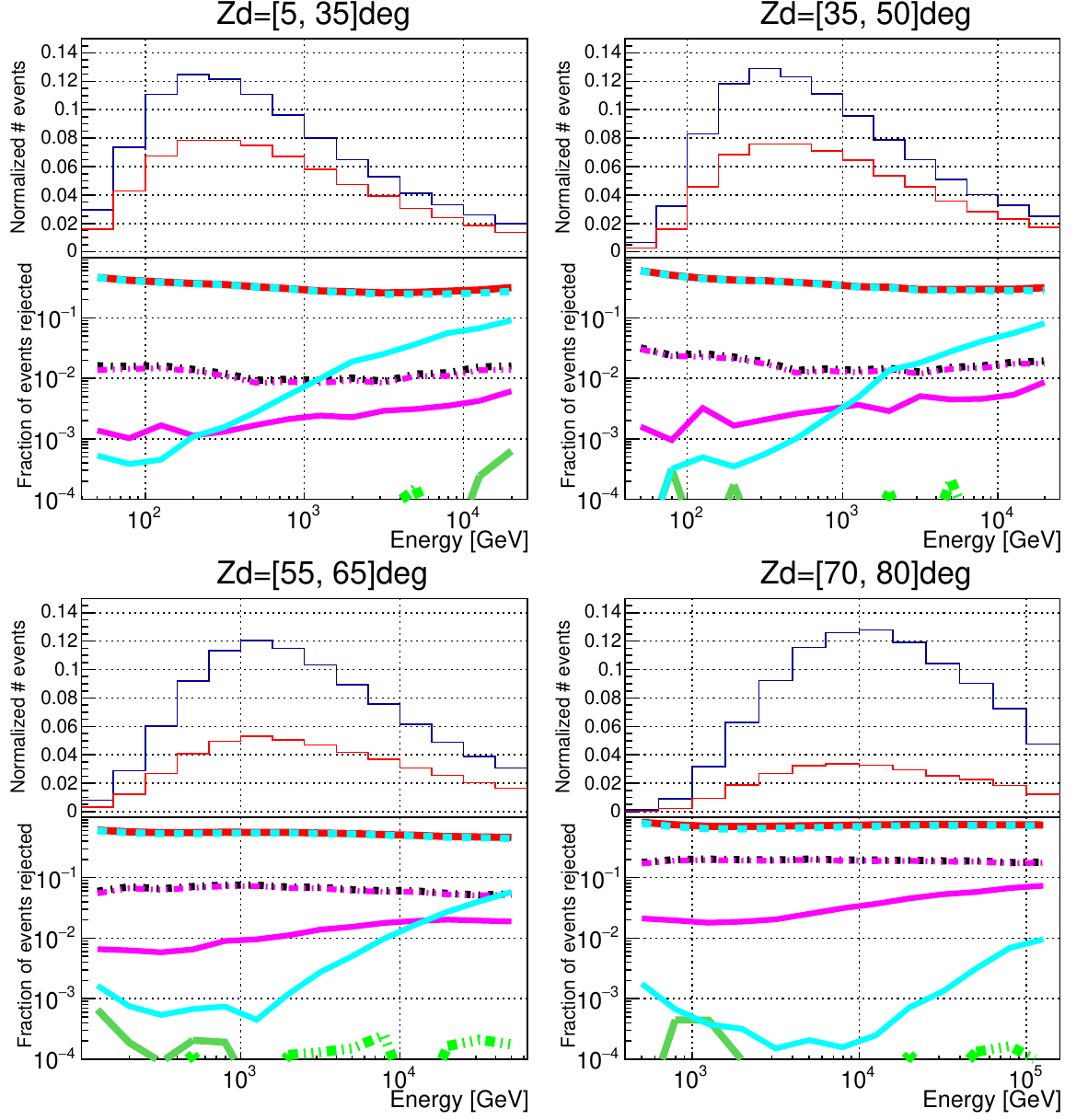}
\caption{Event selection for training \mbox{LUTs-Erec}  in different Zd ranges.
The upper panels show the number of events,  normalized to the total number of MC train samples . The blue and red histograms are the all parametrized events and the used events for training. The fraction of events removed by the individual selection cuts is shown in the lower panels. 
}\label{fig:TrainingEvtsLUT} 
\end{figure}

 \figref{fig:TrainingEvtsLUT}  shows the above-described event selections in different Zd ranges. The cut on Theta2 is responsible for the rejection of several tens of \%, and dominates the overall selection of events used in the training of the LUTs. 

On the other hand, as RF is robust due to the averaging process in calculating the estimand, there is no need to make such a strong event selection before the training.  The only selection cut used in the RF training relates to the rejection of events without a valid reconstruction of the shower axis in the Disp method, that results in the rejection of only a few outliers. The ability of training the RF with events with a wider range of characteristics leads to better performance, especially at large zenith angles and higher energies.

%------------------------------------------------
%  -- Generated by --
% Individual panels:
% magicserv05, 
% /mnt/data/ishio/MAGICana/macro/macro_Eest/EvaluateSurvivedEventsForCutForArticle.C
% combining the four panels: 
% air21
% /Users/kazuma/Workspace/MAGICana/macro_Eest/CombineFourPlotsForLUTtrainingcut.C
%------------------------------------------------

\begin{figure}[t]
\centering
\includegraphics[width=\textwidth]{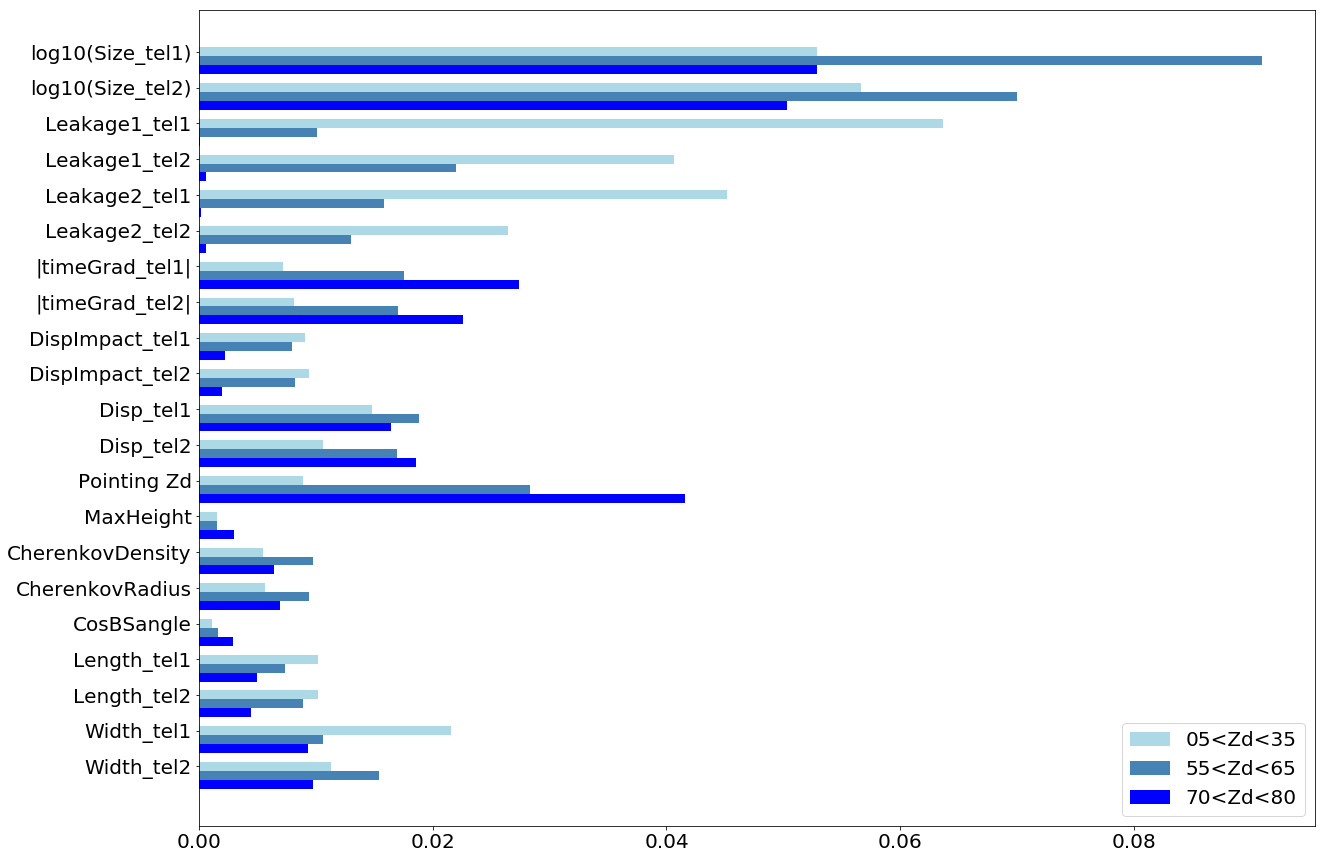}
\caption{Importance scores of the input variables for RF-Erec training.
% \footnotesize
The importance of a variable is calculated as the total reduction in sum of squared errors in the target value, the logarithm of energy.
}\label{fig:Importance} 
\end{figure}

\subsection{Final implementation}
The RF-Erec method was implemented into the standard analysis package developed by the MAGIC collaboration named MAGIC Analysis and Reconstruction Software (MARS) \citep{MAGICanalysis1, MAGIC2016upgrade2}. This software uses C++ routines combined with ROOT libraries \citep{Brun1997}. In the MARS software, RF was originally introduced for the single telescope data \citep{Albert2008_hadronness}. The functionality is developed as C++ adoption of the original Fortran code by Leo Breiman and Adele Cluster \citep{Breiman2001}. 
The RF-Erec was developed based on the library of this functionality. For the split method of this regression problem, we selected the minimization of weighted average of variances in the two sides after split. The determination of the estimand at the last nodes in a tree is the average of the samples at the last node, and the estimand of the entire RF is the average of all the individual trees. The target variable in the RF, the energy, is modified in the form of logarithm base 10, in order to better process the large range of energy values in the train sample, that is described by a power law  extending  over many decades  \citep[see Fig.~4.16 in ][]{Ishio2020}. The main parameters for the forest generation are three: the number of trials for a split, the minimum samples at the last node, and the number of trees. In the RF-Erec that we developed, these parameters are set to 5, 3 and 100, respectively. These values are selected by balancing the computational cost and performance, yet the performance does not change significantly when using other similar values. 

The importance scores of the input variable in the forest generation are shown in  \Figref{fig:Importance}. The MAGIC standard MC productions are used for this study. The scores related to some of the variables have a strong dependency with the zenith angle of the observation. This is expected because the images in the camera change with the zenith angle, and hence the optimal strategy to reconstruct the energy must also change.
%For example, in the lower Zd range, the images of high energy events tend to be larger and sometimes they get truncated, thus the Leakages have greater importance scores, on the other hand, the distance to the shower increases with the Zenith angle which therefore gains greater importance.

\section{Verification}\label{sec:Verification}

Owing to the lack of a calibration source of very high energy gamma rays for an Cherenkov telescope, the validation of the energy reconstruction is done in an indirect manner. In our study, we used three strategies, which are described in the following paragraphs.

\subsection{Data-MC comparison of reconstructed parameters}\label{ParameterDistribution}
%%%%%%%%%%%%%
% motivation
%%%%%%%%%%%%%
The distributions of a number of parameters, e.g., Size parameter, depend on the energy of the incoming gamma ray, and this is the reason for using them in the various methodologies for energy determination. However, the estimator is trained with MC events, not the real events, and hence it is necessary to evaluate if the simulated parameters agree with the real (measured) parameters. 
%For a gamma ray with a given energy coming from a given direction towards a given position on the ground, the measured value of the variables should follow the correspondent probability distribution.
%If the energy reconstruction works consistently in both the simulated gamma rays and the real gamma rays with the same energy spectrum, the measured variables of events should distribute in the same way in a given energy. 
%The energy estimator is based on the parameters of MC data, while it evaluates those of the observation data. Therefore the consistency in the parameter distributions should be seen.  If the MC gamma rays are simulated so that they come from the same direction as a real point-like source and arrive uniformly on the ground, the distribution of the variables should be identical to the real gamma rays with a given energy. In other words, the parameters distribute differently over energy but they should be consistent between the MC and real gamma rays in a given energy. 
%%%%%%%%%%%%%
% real data: ON-OFF
%%%%%%%%%%%%%
A parameter distribution of real gamma rays can be obtained from observation data taken with wobble mode, if there is a high flux of gamma rays from the source, and the gamma-ray signal is comparable or larger than the background. A parameter distribution of the events with their reconstructed incoming directions around the object ("ON" direction) is that of the real gamma rays on top of the background events. The distribution of the background can be subtracted, using the one around the counterpart direction, which is at the same distance from the camera center but in the opposite direction ("OFF" direction). %\footnote{
%This background estimation is based on the instrument response well approximated to radial symmetry. Therefore it is possible to select multiple OFF regions at the same radial distance and area as long as they don't overlap. 
%}. 

%%%%%%%%%%%%%
% real data: selection
%%%%%%%%%%%%%
To this end, we adopted the standard wobble mode observation data of the large flaring activity of the blazar Mrk421 in April 2013, which delivered the highest VHE gamma ray fluxes (during hour and day timescales) recorded by MAGIC to date \citep{Mrk421Apr2013Flare}. The flux was highest during the first four days, April 12, 13, 14 and 15 in 2013. The integrated flux \mbox{$> 200$ GeV} varied from 
\mbox{$1 \cdot 10^{-9}$ cm$^{-2}$s$^{-1}$}  to \mbox{$2 \cdot 10^{-9} $cm$^{-2}$s$^{-1}$}, corresponding to around 5 to 10 times the photon flux of the Crab Nebula. This resulted in $\sim 1.7\cdot10^5$ of excess events in total for this validation study, which amounts to a total observation time of $\sim$ 20 hours, after the selection of the Zd range from 5 deg to 35 deg and very good weather conditions. 
%"wobble mode"\citep{Fomin1994wobble}, with the source direction offset by 0.4 deg from the camera center. 

%%%%%%%%%%%%%
% MC 
%%%%%%%%%%%%%
The MC gamma rays are generated in a generic (universal) manner, which implies that they are with a spectral shape and pointing history that is different from the observation data.
In order to be able to properly compare the MC data with the real data, the former needs to be scaled or reweighted, on individual events, to remove these differences. The weights for normalizing the pointing history are derived from the difference of Zd distributions. The Az direction is not considered because of negligible contribution at low Zd. 
% The figure \ref{fig:ThSqAndZdDist} shows the distribution of $\theta^2$, the angular distance squared, in ON and OFF directions, indicating the substantial amount of the real gamma rays. 
% In the same figure, the adjustment of MC data is shown as well. We adopted the two weights to the MC events. One is for aligning the pointing history, especially the difference of Zd distributions, shown in the top panel of the figure. The Az direction is not considered because of negligible contribution at the low Zd. 
%For the adjustment of the pointing history, the Az direction is not considered because of negligible contribution at the low Zd. Based on the difference of the Zd distributions, the weight is applied to the MC events. For the weight for energy spectrum is also applied. 
As for the weights for scaling the energy spectrum, one needs the true energy spectrum of the real gamma rays. The spectral shape was determined through the spectral analysis on the observation data with the MAGIC standard analysis. The best fit result were obtained with a power law with semi-exponential cutoff, where the forward  folding  method yielded $\chi^2$/n.d.f. = 13.2/16 (p-value = 0.66). The weights were  calculated scaling the measured spectral shape of real gamma rays to that of the simulated MC gamma rays, which is a power law with a differential index of -1.6   %The bottom right panel shows the $\theta^2$ distributions of the observation data after performing ON - OFF, and the MC data after the application of the weights. 

Based on the above mentioned adjustments, the parameter distributions with the highest contributions to energy reconstruction are binned in estimated energy. The distribution separated in estimated energies suffers from deformation due to energy migration, however they should still show the consistency, since the deformation should happen in the same way for both real data and MCs. 

\begin{figure}[t!]
% Size M1
\includegraphics[width=\textwidth]{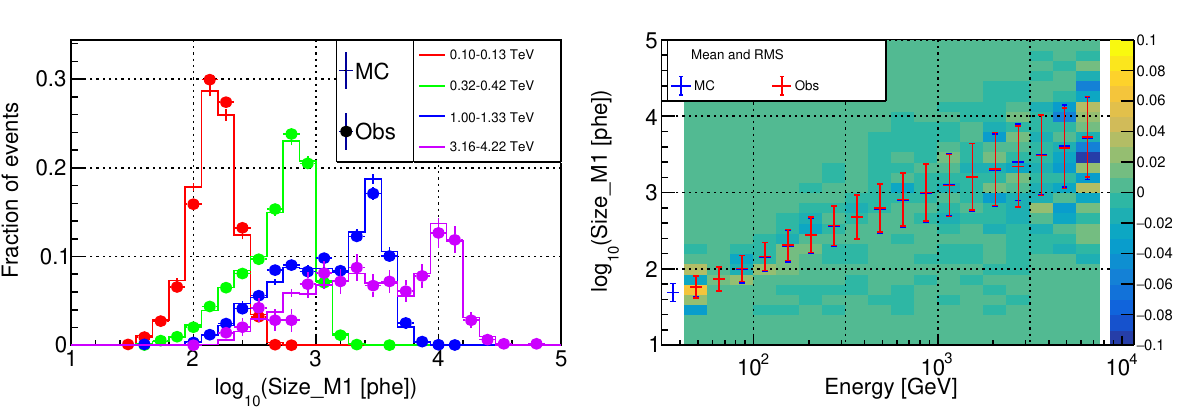}
% Size M2
\includegraphics[width=\textwidth]{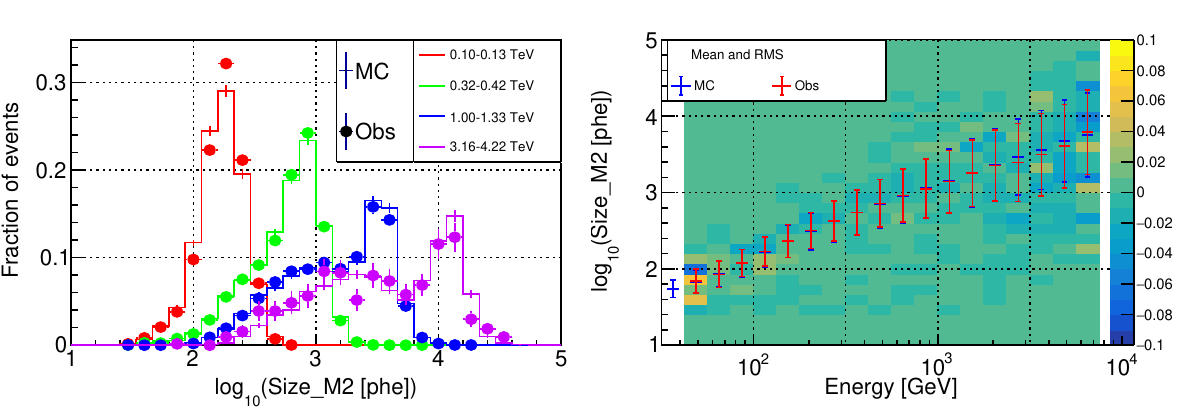}
\caption{
Comparison of simulated gamma rays (MC) and real gamma rays (Obs), obtained as ON-OFF from a dataset with a large signal/background ratio (ﬂare from Mrk421 in April 2013). The left panels show the distributions of Size of the telescope vs estimated energy for MAGIC telescope 1 (M1) in the top and MAGIC telescope 2 (M2) in the bottom. For simplicity, only four selected (representative) energy bands are shown. The MC gamma rays are weighted to match the actual measured gamma-ray spectrum (see main text for details).  The right panels depict the mean and the RMS around the mean in the binned energy ranges, together with the residual of the two histograms normalized by the total number of events in each estimated energy bin of the observation data. Only the energy ranges with more than 100 excess events are presented.
}

\label{fig:MCRealComparisonSize}
\end{figure}

%Compared to the one in true energy, the distribution suffers from the deformation due to the limited energy reconstruction performance, however, since the deformation should happen in the same way on them, they should still show the consistency. 
The results from these comparisons are shown in \figref{fig:MCRealComparisonSize}. %They should show consistent distributions, given the same energy reconstruction is applied to both MC and real data. 
% The parameter distributions are shown in the top left panels of the figure for the MC data, which denote MC gamma rays, and in the top right panels for the real gamma rays as ON - OFF of the distributions. At a glance, the distributions are very similar for all the variables between the MC and the real data. 
%The difference between them is shown in the bottom right panel.The distributions are compared also in the same panel in the bottom left, where four energies are selected. 
% The detailed comparisons of the two distributions are seen in the bottom panels. 
The left panels show the distributions of the parameters in four selected estimated energy ranges, 0.10 - 0.13 TeV, 0.32 - 0.42 TeV, 1.0 - 1.3 TeV and 3.2 - 4.2 TeV. They are separated at equal distances in logarithmic energy to the next ranges, and the width of range is chosen to contain a significant number of events. The shapes are different among the tested energy ranges, yet the MC and the real data agree well with each other. The right panels show the mean and RMS of the distribution in the energy bins, overlaid on top of the residuals of the two dimensional histograms of the parameter values and the estimated energies. In the residual histograms, the distribution of real data is subtracted from that of MC data, and is shown as the ratio to the total number of events in the real data integrated over each energy bin. All the grids with good statistics are within few percent of the total events in the estimated energy bin. %The graph superposed indicates the mean and the RMS of the distribution in each energy.

In this comparison, we see a good agreement of the parameter distributions between the MC and the real gamma rays. In most cases, the difference between the data and the MC is within error bars (or at the level of several \%), but sometimes, for a few bins, the difference could be as big as 10 \%. Those differences are well within the typical systematic errors in Cherenkov telescopes. 

% It is remarkable that the profile histograms mostly agree with each other in different energy ranges, although the shapes of the distributions are complicated and vary in energy. The discrepancies still partially seen would mostly come from the difference of PSF. 

%------------------------------------------------
%  -- Generated by --
% /Users/kazuma/Workspace/MAGICana/macro_CompareParamDist2021/Reshape.C
%------------------------------------------------

\begin{figure}[tp!]
%DispImpact M1
\includegraphics[width=\textwidth]{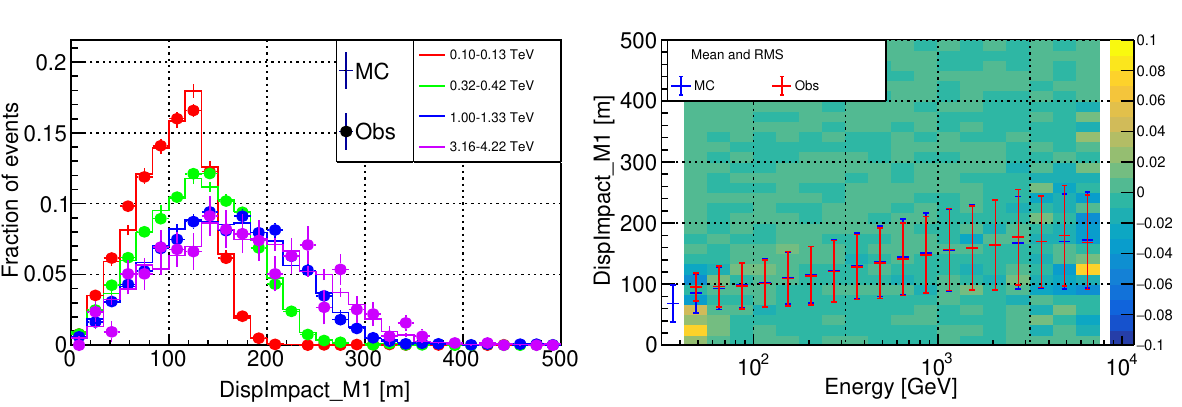}
%DispImpact M2
\includegraphics[width=\textwidth]{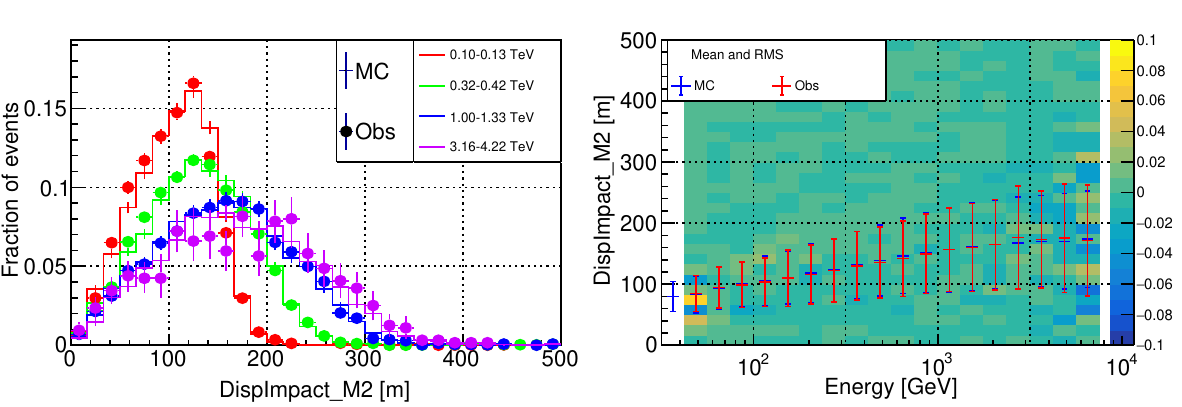}
%Disp M1
\includegraphics[width=\textwidth]{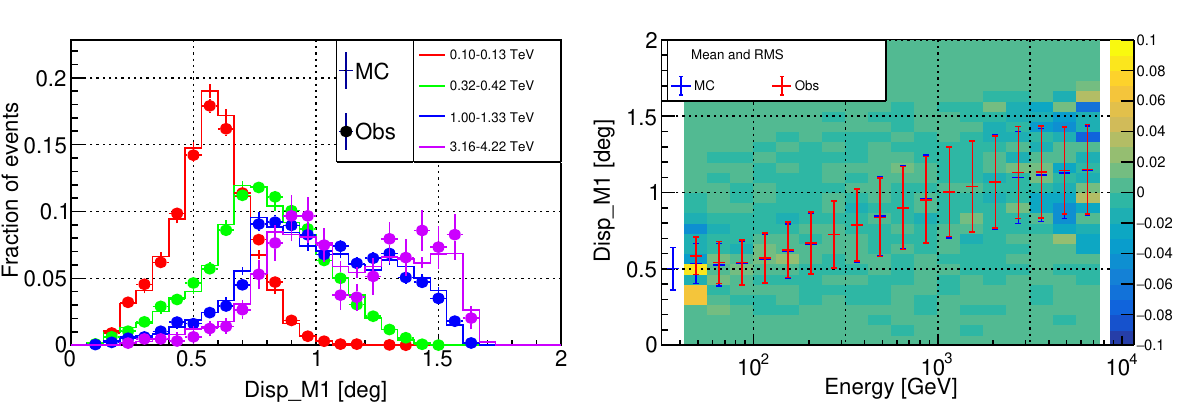}
%Disp M2
\includegraphics[width=\textwidth]{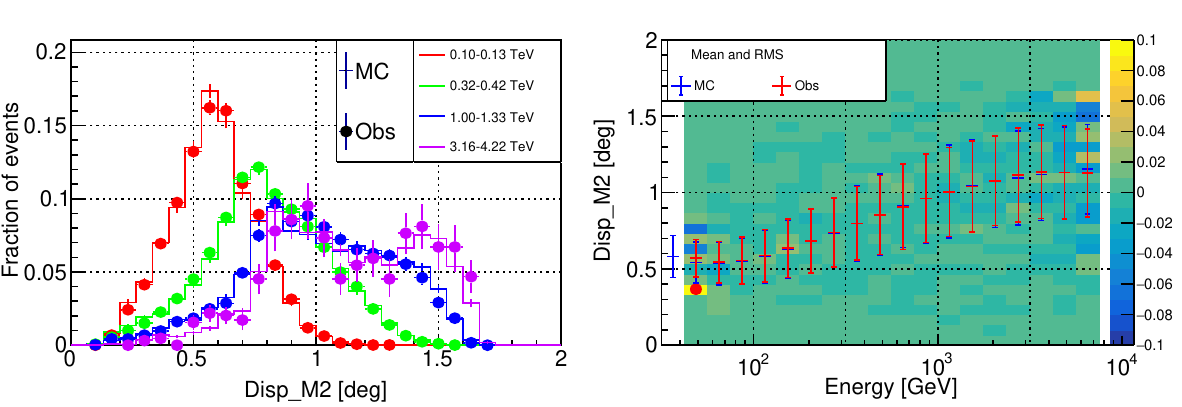}
\caption{Same as \Figref{fig:MCRealComparisonSize}, but for Disp-Impact and Disp.}
\label{fig:MCRealComparisonDisp}
\end{figure}

\begin{figure}[tp]
%Leakage1
\includegraphics[width=\textwidth]{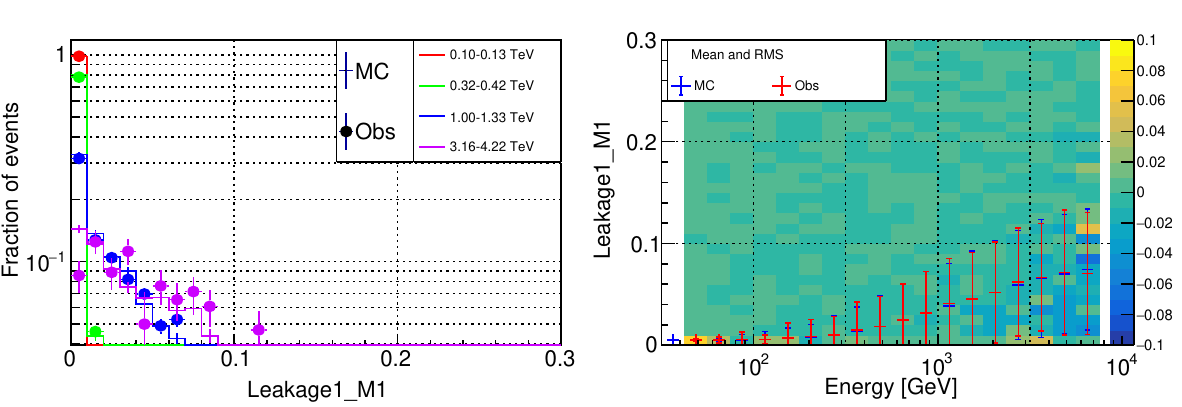}
%Leakage1
\includegraphics[width=\textwidth]{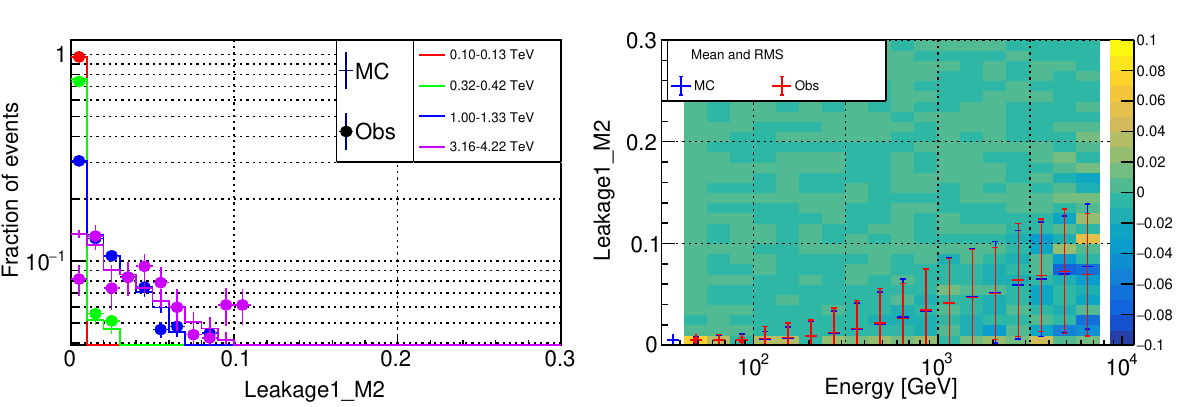}
%Leakage2
\includegraphics[width=\textwidth]{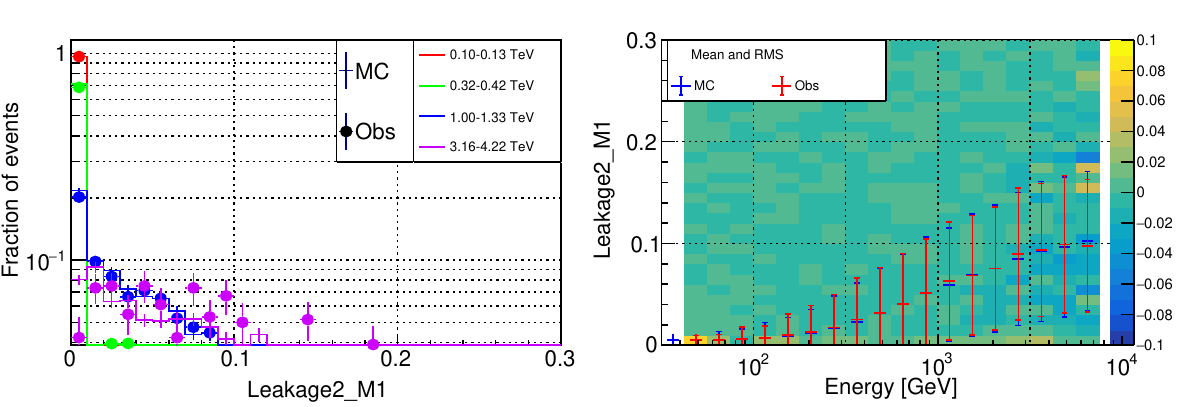}
%Leakage2
\includegraphics[width=\textwidth]{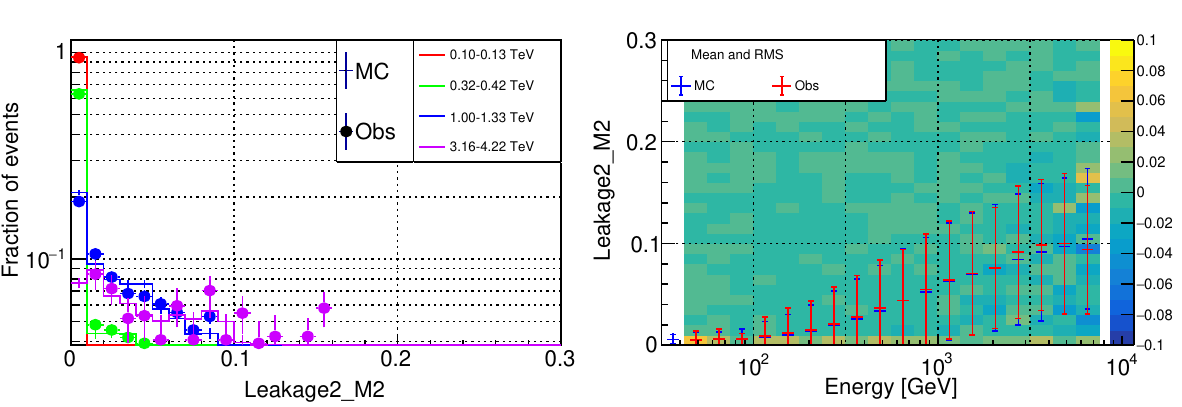}
\caption{Same as \Figref{fig:MCRealComparisonSize}, but for Leakage.}
\label{fig:MCRealComparisonLeakage}
\end{figure}

\subsection{Comparison of the estimated energy from the images recorded in each telescope separately}
The performance of the two telescopes is slightly different because of differences in various telescope elements, such as the mirrors and PMTs \citep{Aleksic2016, MAGICPerformancePaper2012}. 
The MC simulations are naturally adjusted to take into account this inter-telescope differences. The LUTs-Erec consists of the telescope-wise estimators in which the energy is reconstructed from the geometrical parameters and the image parameters of either one telescope.  Among other things, the LUTs-Erec methodology has been validated with consistency between the estimated energies obtained independently from each single telescope, as shown in Fig.~12 in \citet{MAGIC2016upgrade2}. This is very practical consistency test that can be applied to any array of Cherenkov telescopes.

In this manuscript we report the results of this consistency test applied to LUTs-Erec and RF-Erec. Unlike the LUTs-Erec, the RF-Erec is based on single estimator derived with the data from both telescopes. In order to make the single-telescope-wise estimators, like the ones in the LUTs-Erec, the two RF-based estimators were separately constructed by removing the image parameters of either one telescope. 
The results of this consistency test are derived with both, the MC data and the real observation data. To be able to compare the results between the MC and real data, the weights are applied as described in  \secref{ParameterDistribution}.

%------------------------------------------------
%  -- Generated by --
%
% /Users/kazuma/Workspace/MAGICana/Telwise/
%   RF/TelwiseEestDiffQuantile.root
%   LUT/TelwiseEestDiffQuantile.root
% -> 
% /Users/kazuma/Workspace/Python/20220112_Telwise/
%   plotTelwiseDiffQuantile.ipynb
%------------------------------------------------
\begin{figure}[t!]
\includegraphics[width=0.48\textwidth]{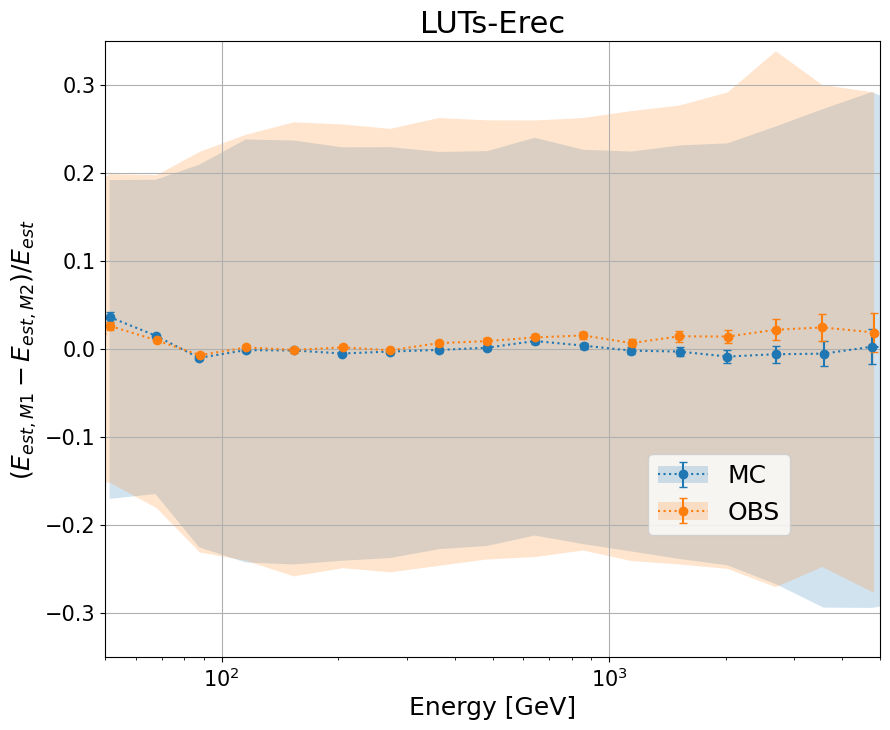}
\includegraphics[width=0.48\textwidth]{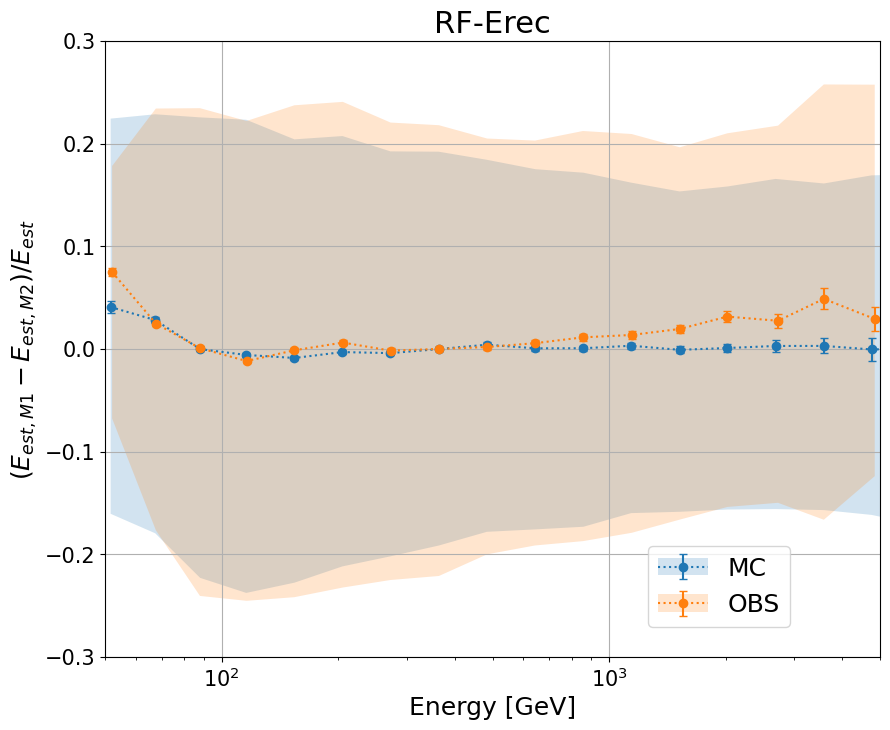}
\caption{%Consistency between two telescopes. 
Comparison of the estimated energy for events from the images recorded in each telescope separately, for both MC and observed gamma-ray data. The plots depict the difference of the single-telescope energies normalised with the stereo-telescope energy. The error bars depict the uncertainty in the determination of the mean, that is retrieved directly from the fit with a Gaussian function. The error bars are hardly visible because, for most energies, their length is comparable or smaller than the size of the markers. The shaded regions show the 68\% containment of the distributions for each energy bin. The left panel reports the results for LUTs-Erec, and the right panel for the RF-Erec.  
}\label{fig:TelwiseDiff}%\label{fAccuracy}
% \begin{quote}
% \end{quote}
\end{figure}

The results of this consistency test are depicted in \figref{fig:TelwiseDiff}, separately for LUTs-ERec and RF-Erec. The distributions of differences in the single-telescope energies are centered around zero, indicating that the inter-telescope calibration in the MC data is successfully done. In both RF-Erec and LUTs-Erec, the mean and the 68 \% containment are consistent between the MC and real data within a few percent accuracy.
These differences are sufficiently small considering that the overall systematic uncertainty in the energy determination is in the order of 15\%  \citep[see section 4.4 of ][]{MAGIC2016upgrade2}

\subsection{Validation of RF-Erec with the energy spectrum of the Crab Nebula}\label{sec:CrabCheck}
A very common method to validate the different energy reconstruction methods is by means of the evaluation of the reconstructed gamma-ray spectrum for Crab Nebula, the "standard candle" in the VHE gamma-ray astronomy.
We evaluated the measured Crab Nebula spectra obtained with data in four Zenith distance ranges:  Zd=[5,35]deg (LZd), Zd=[35,50]deg (MZd), Zd=[55,65]deg (HZd), and Zd=[70,80]deg (VHZd). All the observations were performed during astronomical night, with no moon in the sky or near the horizon, and during good atmospheric conditions, in order to minimise the systematic uncertainties.
The total exposure of the data used in this evaluation test is 11.5 hours, 5.8 hours, 3.7 hours and 20.5 hours for LZd, MZd, HZd and VHZd, respectively.
%\footnote{
%There are different versions of the MC data available in the standard production, representing specific periods. A period is defined based on different telescope performance due to a significant performance change by weather conditions, minor hardware maintenance or major hardware intervention.
%In this paper, all the MC data selected are compatible with the telescope performance to date in 2021.
% All the selected data are in the period from 2016-04-29 to 2017-08-02. This period is compatible with the telescope performance to date in 2021, as there has been no major hardware intervention or a significant performance change due to weather conditions or minor hardware maintenance.
%} .

%The flux points are calculated by the number of excess events divided by collection area, which was corrected for the energy migration using the migration matrix\footnote{
%The migration matrices used are presented in \figref{fig:MigMatrix}. 
%}, collection area for true energy and the assumed spectrum obtained from the past study \citep{Albert2008a}. The cut condition is 
%dependent on estimated energy 
The gamma-ray spectra were derived following the MAGIC standard analysis, that includes the event selection. For all the bins in estimated energy, the selection applied retains the events with a Size above 50 photoelectrons that survive the angular distance and the hadronness cuts that have a gamma efficiency of 75\% and 90\%, respectively \citep{Albert2008_hadronness}. 
The number of excess events and background events, from the ON region and three OFF regions available by the wobble pointing, are fitted using the forward folding method with a curved power-law fit function:
\begin{equation}
\diff{N}{E}=  f_{0}\cdot (E/1\,\mathrm{TeV})^{a + b \log_{10} (E/1\,\mathrm{TeV})} \rm [cm^{-2} s^{-1} TeV^{-1}]\, , 
\end{equation}
where $f_{0}$, $a$ and $b$ are the amplitude, the power-law index and the curvature. This function yields a good description (p-value$>$0.1) of the Crab Nebula gamma-ray spectra for all the Zd ranges, and for both \mbox{LUTs-Erec} and \mbox{RF-Erec}.

%------------------------------------------------
%  -- Generated by --
% /Users/kazuma/Workspace/Python/20220808_CrabSEDs/CrabSEDPlotter.py
%------------------------------------------------
\begin{figure}[htbp]
\includegraphics[width=\textwidth]{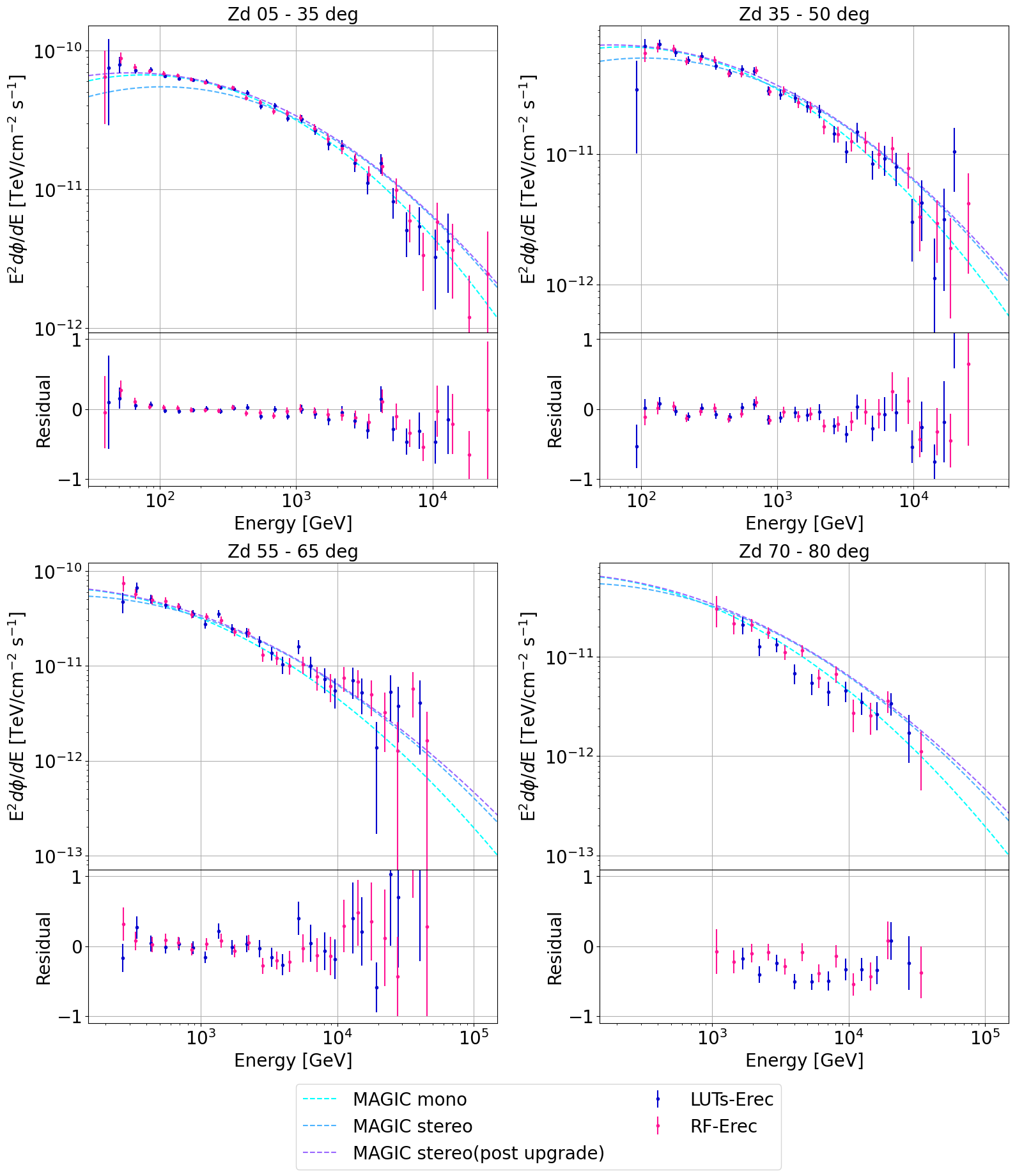}
\caption{ %Validation of the energy reconstruction via the Crab Nebula spectra. 
The Crab Nebula spectra obtained with \mbox{LUTs-Erec} and \mbox{RF-Erec} in different Zd ranges. The points are calculated taking into account the energy migration based on the fitted spectral shape by forward folding. The lines are the spectra from the past publications by the MAGIC collaboration, indicated in the legend as "mono"\citep{Albert2008a}, "stereo"\citep{MAGICstereoPreUpgrade} and "stereo (post upgrade)"\citep{MAGIC2016upgrade2}. The lower panels show the normalized residual from the latest MAGIC publication, the "stereo (post upgrade)". 
}\label{fig:CrabCheck}
% \caption{ %Validation of the energy reconstruction via the Crab Nebula spectra. 
% The Crab Nebula spectra obtained with \mbox{LUTs-Erec} and \mbox{RF-Erec} are shown in blue and magenta respectively, in different Zd ranges. The dashed lines are the spectral functions by the forward unfolding fit, and the points are calculated taking into account the energy migration and the fitted spectral shape in the collection area derived in estimated energy. The spectrum from the latest study with MAGIC \citep{MAGIC2016upgrade2} is drawn together. The lower panels show the normalized residual from the latest study. 
% }\label{fig:CrabCheck}
% \begin{quote}
% \footnotesize
% % From top to bottom, the Zenith distance range of the observations are indicated in the panels. The red dashed line is the Crab Nebula spectrum presented in past study by MAGIC collaboration % \cite{Albert2008a}.
% \end{quote}
\end{figure}

The results from the above-mentioned spectral fits are reported in \Figref{fig:CrabCheck},  in the form of spectral energy distribution, i.e. $E^2 dN/dE $. The data points are calculated from the result of forward folding, namely, taking into account the energy migration and the fitted spectral shape in the collection area derived in estimated energy.
%They are calculated taking into account the energy migration and the fitted spectral shape in the collection area derived in estimated energy. 
As a reference, the panels also show three Crab Nebula spectra previously published by the MAGIC collaboration (at different stages of the instrument): \cite{Albert2008a, MAGICstereoPreUpgrade, MAGIC2016upgrade2}. In order to better quantify the agreement of the recent spectral measurements (obtained with LUTs-Erec and RF-Erec) with those published, in the lower panels, we also depict the residual of the spectral points with respect to the last Crab Nebula spectra published by MAGIC, namely \citet{MAGIC2016upgrade2}. Considering that the systematic uncertainties in Cherenkov Telescopes, when taking into account both the absolute energy scale and the flux scale, are at the level of 20 -- 30\%    \citep{MAGICstereoPreUpgrade, MAGIC2016upgrade2, TheHESSCollaboration2006},  we conclude that the data points agree very well with the reference spectra. In the highest Zd range, VHZd, the deviations with the reference spectra are somewhat larger than at the lower Zd ranges (specially for the \mbox{LUTs-Erec}). However, the differences are within the expectations, owing to the larger systematic uncertainties related to such high Zd observations, as reported in \citet{Acciari2020c}.

% CrabNebula_ST0307_Zd05-35_RFv5/Thsq75_Had90_40bins
% Final Chi2: 22.8531 / 24
% Prob = 0.528486  

% CrabNebula_ST0307_Zd05-35_LUT/Thsq75_Had90_40bins
% Final Chi2: 21.2028 / 23
% Prob = 0.568705   

% CrabNebula_ST0307_Zd35-50_RFv5/Thsq75_Had90_40bins_Az12bin
% Final Chi2: 30.5169 / 22
% Prob = 0.106432            

% CrabNebula_ST0307_Zd35-50_LUT/Thsq75_Had90_40bins_Az12bin
% Final Chi2: 28.6291 / 23
% Prob = 0.192922      

% CrabNebula_ST0307_Zd55-65_RFv5/Thsq75_Had90_40bins_Az12bins
% Final Chi2: 25.9487 / 20
% Prob = 0.167513    

% CrabNebula_ST0307_Zd55-65_LUT/Thsq75_Had90_40bins_Az12bins
% Final Chi2: 24.0713 / 20
% Prob = 0.23929          

% CrabNebula_ST0307_Zd70-80_RFv5/Thsq75_Had90_40bins_Az12bin
% Final Chi2: 11.2346 / 11 
% Prob = 0.423827      

% CrabNebula_ST0307_Zd70-80_LUT/Thsq75_Had90_40bins_Az12bins/
% Final Chi2: 9.84651 / 10                                                                           
% Prob = 0.454061               

\section{Performance}\label{sec:Performance}

In this section, we compare the performance of the new energy estimator, \mbox{RF-Erec}, to that of \mbox{LUTs-Erec}, which is the method that has been used by the MAGIC collaboration during more than one decade. This section has four subsections where we discuss the migration matrices, the energy resolution and bias for point-like and extended (or off-axis point-like) sources, and we also evaluate the distortions in the spectra produced by the migration of events.

\subsection{Migration matrix}\label{sec:migmat}
The migration matrix contains the probabilities for given true energies of gamma rays to be reconstructed into a given bin in estimated energy. It is constructed based on the test samples, namely an additional set of MC gamma rays dedicated for evaluations of the estimators and the instrument response function. To this end, the test samples are processed in the same event reconstruction and event selection as the standard analysis. 
Also, the weight for Zd and energy distribution are applied to depict the specific Crab Nebula observation data. 

%------------------------------------------------
%  -- Generated by --
% /Users/kazuma/Workspace/MAGICana/macro_Eest/macro_MigMatSED2/ExtractMigMatForPaper_Eff75.C
%------------------------------------------------
\begin{figure}[htbp]
\includegraphics[width=0.9\textwidth]{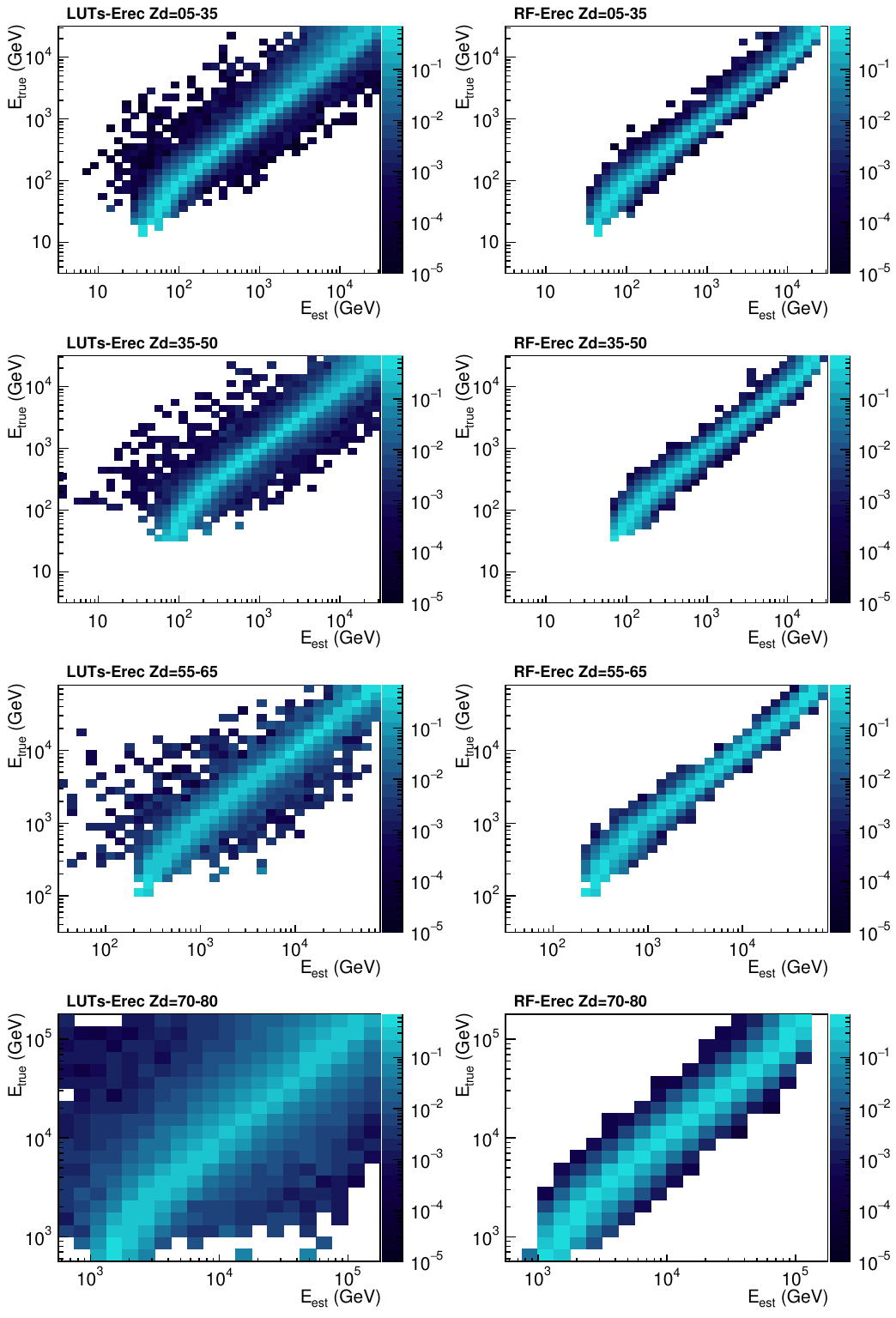}
\caption{Migration matrices from \mbox{LUTs-Erec} and \mbox{RF-Erec} in different Zd ranges.
}\label{fig:MigMatrix} 
\end{figure}

The result is shown in \figref{fig:MigMatrix}, for the four Zd ranges and for both LUTs-Erec and RF-Erec. One can see that
the migration matrices obtained with \mbox{LUTs-Erec} show a wide spread of migration probabilities while the ones from \mbox{RF-Erec} have a tight band of bins along the diagonal line, indicating a better energy resolution. We also note that the spread in the migration matrix of LUTs-Erec, relative to that of RF-Erec, is larger as we increase the zenith angle range, and hence one expects that the performance improvement provided by RF-Erec increases with the zenith angle of the observation. 

Besides the overall shape, in the \mbox{LUTs-Erec} migration matrices one can see a few isolated cells, that could introduce features when reconstructing the spectra. For instance, in the \mbox{LUTs-Erec} migration matrix from \mbox{Zd=[55,65] deg}, one can see isolated cell with events of true energy around 200 GeV with a 5 $\%$ probability to migrate to an estimated energy of about 5 TeV (i.e., 20 times higher energy). We investigated these features, and learnt that they are caused by the usage of "True" MC values in the training sample, such as the usage of True-Impact, in a region of the phase space with limited statistics, such as close to the energy threshold, where only a few events survive, and particularly for high zenith distances because the showers are farther away from the telescopes, and the error in the estimation of the impact is larger. As mentioned in section\ref{sec:true-vs-recon}, the RF-Erec method does not use "True" MC values, only "reconstructed ones", and hence it does not show these artifacts in the migration matrix. Ultimately, this leads to a more reliable determination of the energy spectra in the gamma-ray sources.

\subsection{Generalization error}
In the context of evaluating an estimator, the generalization error is a measure of how accurately an algorithm is able to predict outcome values for previously unseen data.
We evaluate here the generalization error of energy from two perspectives: the energy bias and the energy resolution.
The bias is the averaged difference of the prediction of our model from the correct value, and the resolution is the spread of the model prediction. These general trends are investigated for the distributions of the normalised error, $(E_{est} - E_{true})/E_{true}$ in the test samples, divided into bins of true energy \footnote{
The events are selected in the same way as the standard analysis. Note that the energy resolution can be easily improved in the multi-TeV range with tighter quality cuts, at the price of lowering the collection area.
}. 
As it is widely done in the community, the bias and the resolution are obtained through a Gaussian fit to the above-mentioned distribution of normalized errors
\footnote{
The fit is applied to the samples after standard cuts, and the fit range is determined from the average and the standard deviation (SD) of the population, where the lowest/highest edge of the range is the average $+/-$ the SD. 
}: the bias and the resolution are the peak position and the sigma ($\sigma$) of the fitted Gaussian, respectively. 

However, the energy resolution computed in this way does not take into account the tails of the distribution of normalized errors, which show up very clearly in the migration matrix (outside the central part, containing the 68\% of the events), and play an important role in the energy reconstruction. In order to tackle the problem in a more generic manner, in the quantification of the performance that we report here, we use two energy resolutions: the widely used or standard resolution, which here we name {\it resolution-$\sigma$}, and the standard deviation (SD) of the distribution of normalized errors, which we name {\it resolution-SD}. The latter one is more informative about the shape of the distribution outside the central region.

\begin{figure}[htbp]
\centering
\includegraphics[width=0.7\textwidth,clip,trim=0 0 0 0.45cm]
{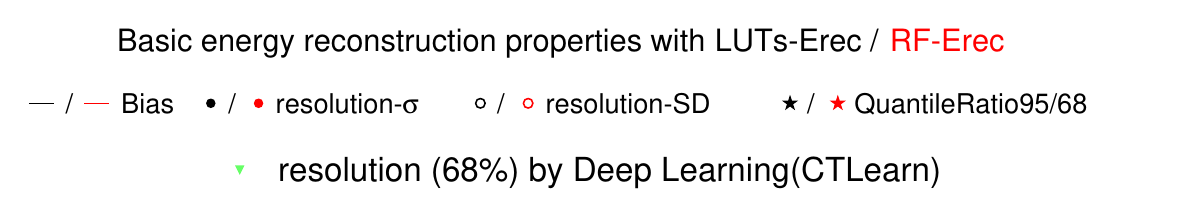}
\includegraphics[width=0.97\textwidth,clip,trim=0 0.3cm 0 0]
% {EvaluateMultiEest_RFv5chDispDt_ST0307_AllZd}
{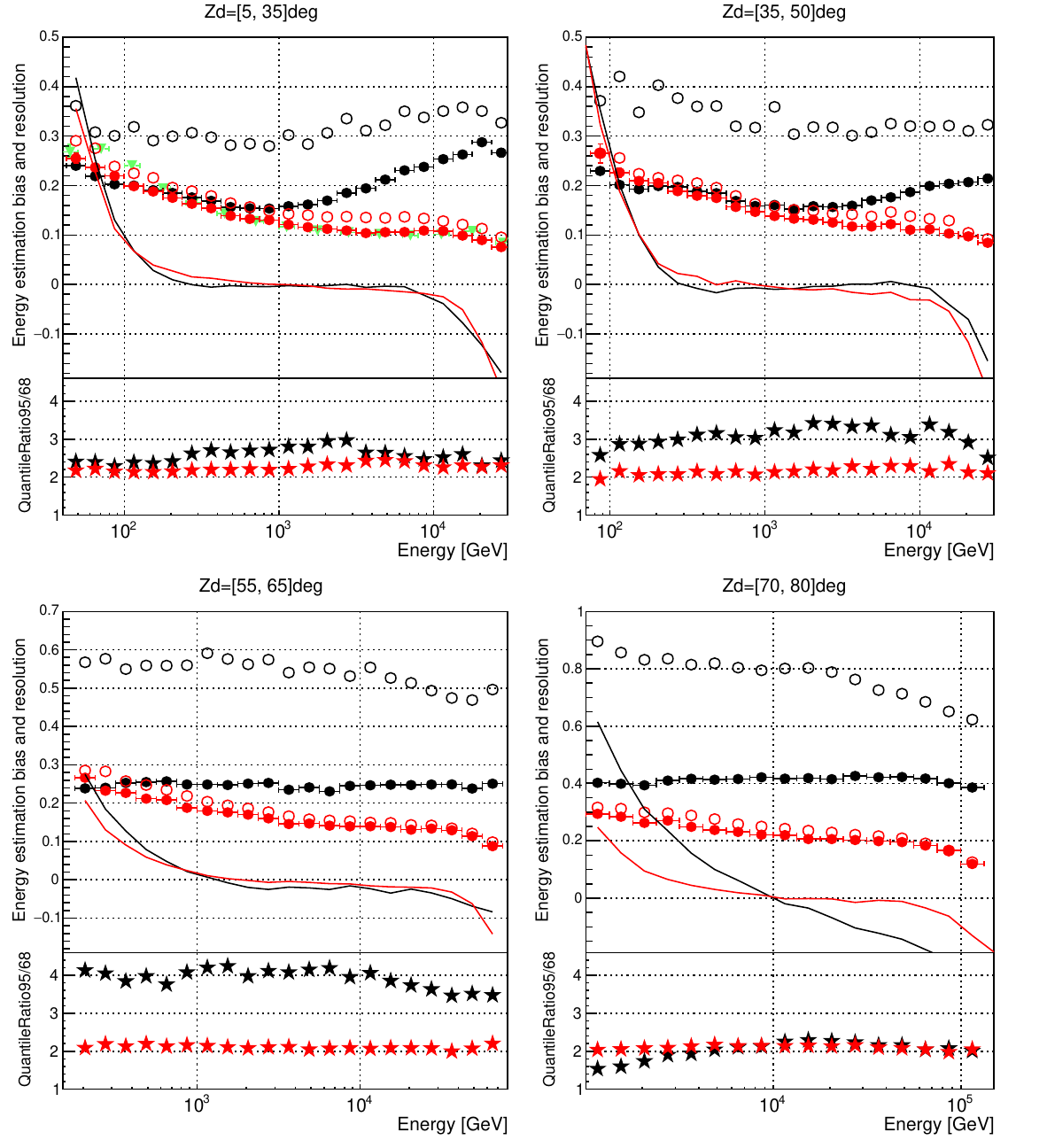}
\caption{%Basic performance of \mbox{LUTs-Erec} and \mbox{RF-Erec} in different Zd ranges
Performance of the \mbox{LUTs-Erec} (black) and \mbox{RF-Erec} (red) energy estimators shown as a function of true energy, separately for four different ranges of Zenith distance. %The distributions of $(E_{est}- E_{true})/E_{true}$ in the test samples are evaluated. \\
The top subpanels show the bias and the resolutions. The bias and the resolution-$\sigma$ are derived from the Gaussian fit, while the resolution-SD is the standard deviation of the population (see main text for further details). 
The resolution of Deep Learning study\citep{Miener2021} is also overlaid. 
%For more details, see \secref{sec:EvaluationDefinition}.
The bottom subpanels 
%are the quantile ratio, which is 
show the ratio of the quantile widths of 95\% containment around the peak of the distribution to that of 68\%, which quantifies of how big are the tails of the distributions. In a Gaussian function, this quantile ratio is 2.0.
}\label{fig:GeneralizationErr}
% \begin{quote}
% \footnotesize
% \end{quote}
\end{figure}

The upper panels of \figref{fig:GeneralizationErr} report the performance of both, \mbox{LUTs-Erec} and \mbox{RF-Erec}, evaluated separately in four Zd ranges.  One can see that, for both methods, the resolution worsens at low energies, and that the bias is largest at the lowest and the highest energies.  The degraded resolution at low energies is due to the dimmer and smaller shower images, which affect the reconstruction by higher fluctuations in estimating the total light content and the distance to the shower. The bias rapidly increases at low energies due to the threshold effect, because the events are detected when there is an up-fluctuation in the amount of light measured by the telescopes. There is a negative bias (energy is underestimated) also at the highest energies, that is visible in both the LUT and RF methods for all the zenith angles. This is caused by the limited energy range in the MC simulations and by other effects related to the limited FoV in the camera. The stronger negative bias is seen in the low zenith angle observation, presumably due to a fraction of the shower images being outside the camera, sometimes even missing the shower maximum, and hence not well reconstructed. As we focus on the RF energy reconstruction method itself, we leave this as out of scope of this manuscript.

% The negative bias (energy is underestimated) at the highest energies is caused by the limited energy range of the MC simulations, which prevents the method to be trained with up-fluctuations, and only with down-fluctuations. We note that the maximum energy of the MC simulations also leads to an artificial (not real) decrease in the energy resolutions. But this effect is very small, and not as visible as the effect in the bias. One could naturally solve these artifacts by increasing the maximum energy in the MC simulations, at the cost of higher computation time. But we did not do that because we consider this is not a critical problem for the evaluation of the performance we report here.

As shown in \figref{fig:GeneralizationErr}, the performance of \mbox{RF-Erec} is substantially better than that of \mbox{LUTs-Erec} over a large range of energies, and in all four Zd ranges. As indicated by the migration matrices (see section \ref{sec:migmat}), the differences between \mbox{RF-Erec} and \mbox{LUTs-Erec} increase with the Zenith angle of the observation: the higher the zenith angle, the larger superiority in the performance of RF-Erec with respect to that of LUTs-Erec.

 In the low Zd range, the improvement of resolution-$\sigma$ is noticeable above 1 TeV, where the resolution-$\sigma$ from \mbox{RF-Erec} reaches around 10 \% at a few TeV, while the one from \mbox{LUTs-Erec} never goes below $\sim 15$ \%, and becomes worse ($\sim 30$ \%) as energy increases. 
 %In the low energy side, the resolution ($\sigma$) are comparable at mostly $\sim 20 \%$. 
Moreover, the resolution-SD from \mbox{RF-Erec} outperforms that from \mbox{LUTs-Erec} for most energies, by a factor of two in the TeV range.

For RF-Erec, the resolution-$\sigma$ and resolution-SD are very similar, indicating that the distribution of normalized errors is close to a Gaussian function, even far away from the central region. On the other hand, the large discrepancy between the resolution-$\sigma$ and the resolution-SD in the \mbox{LUTs-Erec} indicates large deviations from a Gaussian distribution, and hence the presence of large amount of outliers that worsen the measurement of the energy spectra of the gamma-ray sources. 

Another manner of quantifying the importance of the tails in the normalized error distributions obtained with RF-Erec and LUTs-Erec is through the quantile ratio 95/68, which is the ratio of the quantile widths of 95 \% containment around the peak of the distribution to that of 68 \%. For a Gaussian distribution, this value is 2. We report this quantity in the lower panels of \figref{fig:GeneralizationErr}. In the case of \mbox{RF-Erec}, the quantile ratio is always around 2, further confirming the good behaviour of this energy reconstruction method. On the other hand,  in the case of \mbox{LUTs-Erec}, this value is always substantially larger than 2, specially as we increase the Zd range (it reaches 4 in Zd range from 55 to 65). We note that, in the Zd range from 70 to 80, the quantile ratio 95/68 from \mbox{LUTs-Erec} is small (sometimes even below 2) because the resolution-SD is already very large (larger than 0.8, $\sim$3 times larger than that for RF-Erec), and the distribution is truncated by the limited energy range in the MC simulations used in the training.

\subsection{Dependency of the generalization error with the Off-axis angle}

Many of the gamma-ray sources studied with Cherenkov Telescopes have an extension of around 1 degree (e.g., Supernova Remnants), and hence it is important to quantify the performance of an energy reconstruction method for different regions of the field of view.  In this section we report the evaluation of the generalization error for the events from the outer region in the field of view. We used one of the standard MC productions for this purpose, where the incoming directions of MC gamma rays are homogeneously distributed in a region with 1.5deg angular diameter around the center of the field of view. 
%The majority of possible events to trigger is sufficiently included in this region, as the trigger region is 1 deg in the central region of the camera. 
The data set was produced with a pointing direction between 5 and 35 degrees in Zenith angle.

The data set was divided into train and test samples, and used for the construction of estimators and their evaluations respectively. 
% \footnote{
% %  in the period ST0306, 20141122 20160428
% The available data is in the period from 2014-11-22 to 2016-04-28. This period is compatible with the telescope performance to date in 2021, as there has been no major hardware intervention or a significant performance change due to weather conditions or minor hardware maintenance.
% }. 
The evaluation is done for three regions of the test samples' reconstructed incoming directions: below 0.4 degree, from 0.6 to 0.8 degree, and above 1.0 degree from the center of the camera.

\begin{figure}[htbp]
\centering
\includegraphics[width=0.8\textwidth,clip,trim=0 1cm 0 0]
{EvaluateMultiEest_RFv5chDispDt_ST0307_Zd5-35_wDL_legend}
\includegraphics[width=\textwidth]{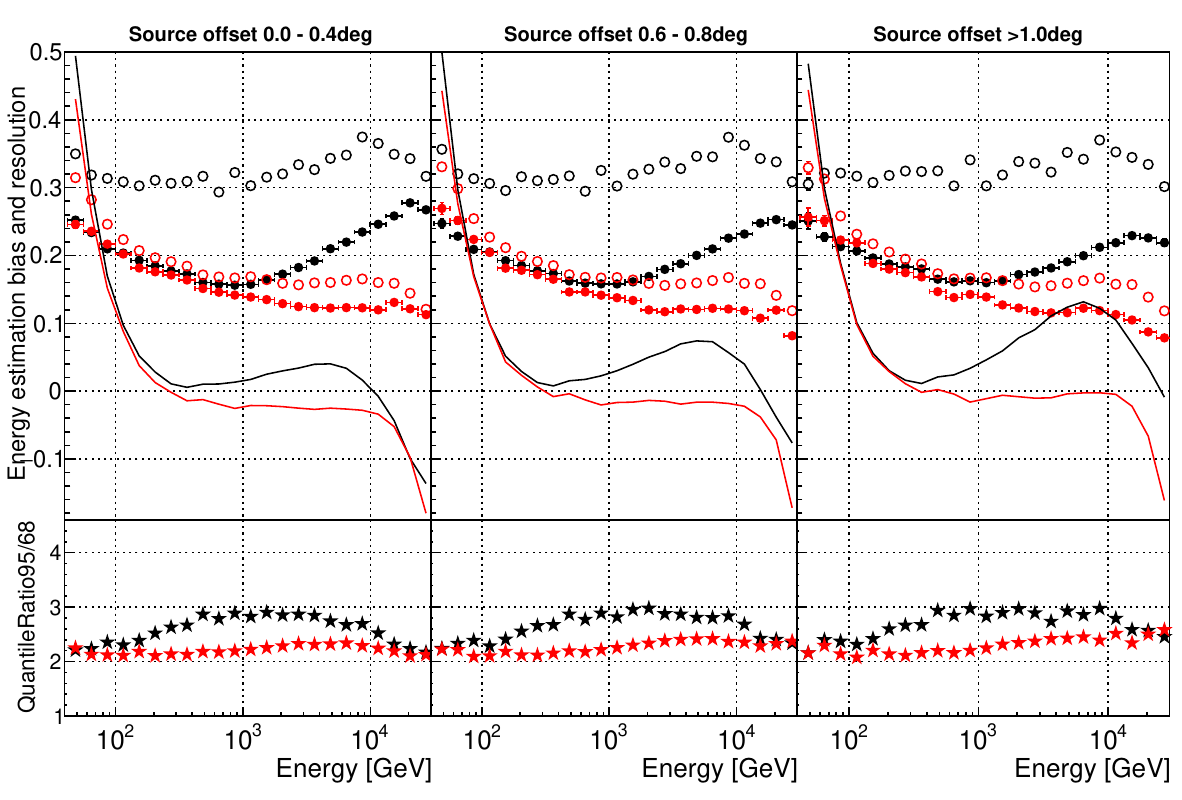}
\caption{Basic performance of \mbox{LUTs-Erec} and \mbox{RF-Erec} for different offsets with respect to the center of the field of view.}\label{fig:GeneralizationErrMorphology}%\label{fAccuracy} 
\end{figure}

\figref{fig:GeneralizationErrMorphology} shows the performances for the three offset ranges. Overall, the performance of both energy estimators is similar for the three offset angle ranges investigated, and comparable to those shown in the top left panel from \figref{fig:GeneralizationErr}. However, one can see that the energy bias from LUTs-Erec shows a large increase at TeV energies, which increases with the offset angle. This increase in the bias is due to a systematic deviation in the correction formulae used in the LUTs-Erec method, which was initially developed only for the standard offset angle.

%------------------------------------------------
%  -- Generated by --
% 20211217
% Created by astro, MAGICana/macro/macro_Eest/EvaluateMultiEest_wCutConfig_quantile.C
% -> MAGICana/macro/macro_Eest/DispImp_Disp_LW/*
% /Users/kazuma/Workspace/MAGICana/macro_Eest/20211217_wEffCut_quantile_diffuse/
%    diffuse_00-04.root 
%    diffuse_06-08.root 
%    diffuse_10.root 
% -> CombineThreePlots.C
%2022/08/27
% /Users/kazuma/Workspace/20220827_EvaluateEest_diffuse/diffuse_three_offsets.pdf
%------------------------------------------------

\subsection{Distortions in the reconstructed gamma-ray spectra due to the migration of events}

The determination of the energy has an uncertainty which naturally yields to differences in the distribution of estimated energies, in comparison to the distribution of true energies for a given population of MC simulated events. 
Because of the power-law-like nature of the spectra of the typical gamma-ray sources \footnote{
The energy distribution of observed events is also power-law-like because of relatively flat collection area far from the threshold. In such distribution, lower energy events are always more abundant than in the higher energy ones, thereby the tiny fraction of spillover to an estimated energy bin from the lower energy side results in a considerable amount. Such contamination can overwhelm the genuine events and cause a systematic difference of the slope in the reconstructed energy spectrum.}, the largest deformation is produced when the estimated energy is larger than the true energy, that is a migration of events from low energies (where there is a big number of events) to high energies (where there is a small number of events). This deformation needs to be corrected using the migration matrix via either forward folding or unfolding \citep{Albert2007}. However the correction leaves uncertainty from the ambiguity given by the irreducible background events as well as by the limited precision in producing the migration matrix from MC data. This ambiguity is the major drawback in a spectral analysis and we assess it by quantifying the reliability of a bin in the estimated energy distribution. %We discuss the reliability in two ways, the bin purity and the spillover ratio. 
%The assessment requires a selection of a pointing direction, which changes the performance of the energy estimator, and also a spectral shape, whose spectral slope significantly changes the influence to the higher energy bins. 
In order to quantify these effects, we select two different ranges of the pointing direction and two slopes of spectrum; low/high Zd (Zd ranges of from 5 to 35 degrees and 55 to 65 degrees), and  hard/soft spectrum (spectral indices of 2.5/ 5.0). Based on these selected Zd ranges and spectra, the relations between the true energy and estimated energy distributions are generated in the following way.

For a given Zd range, an example pointing history of observation is determined, which is used for generating the collection area and the migration matrix from the MC events. For a given source with the given observation, the expected true energy distribution per unit time of the acquired gamma-ray events can be expressed as the source spectrum multiplied by the collection area. The expected distribution over the estimated energies per unit time is obtained by multiplying the energy migration matrix to the true energy distribution. The errors are calculated from those of the collection area and the migration matrix. We decided to use 7 bins per decade in both $E_{est}$ to that of $E_{true}$, so that they can be compared bin by bin and a bin should contain a sufficient quantile in the range of estimated energy for the median energy event assuming a resolution of $\sim 15 \% $.
%\footnote{
%For the steep source, the evaluation was done separately for low energy and high energy, because the number of events are enormously different. The low energy and high energy evaluation has the overlap region so that the purity continues smoothly.}. 
Based on them, we discuss the performance from two perspectives: the bin purity and the spillover ratio.

\subsubsection{Bin purity}\label{sec:BinPurity}
\begin{figure}[htbp]
\includegraphics[width=0.5\linewidth]%,pagebox=cropbox,clip,trim=0 0 0 0]
{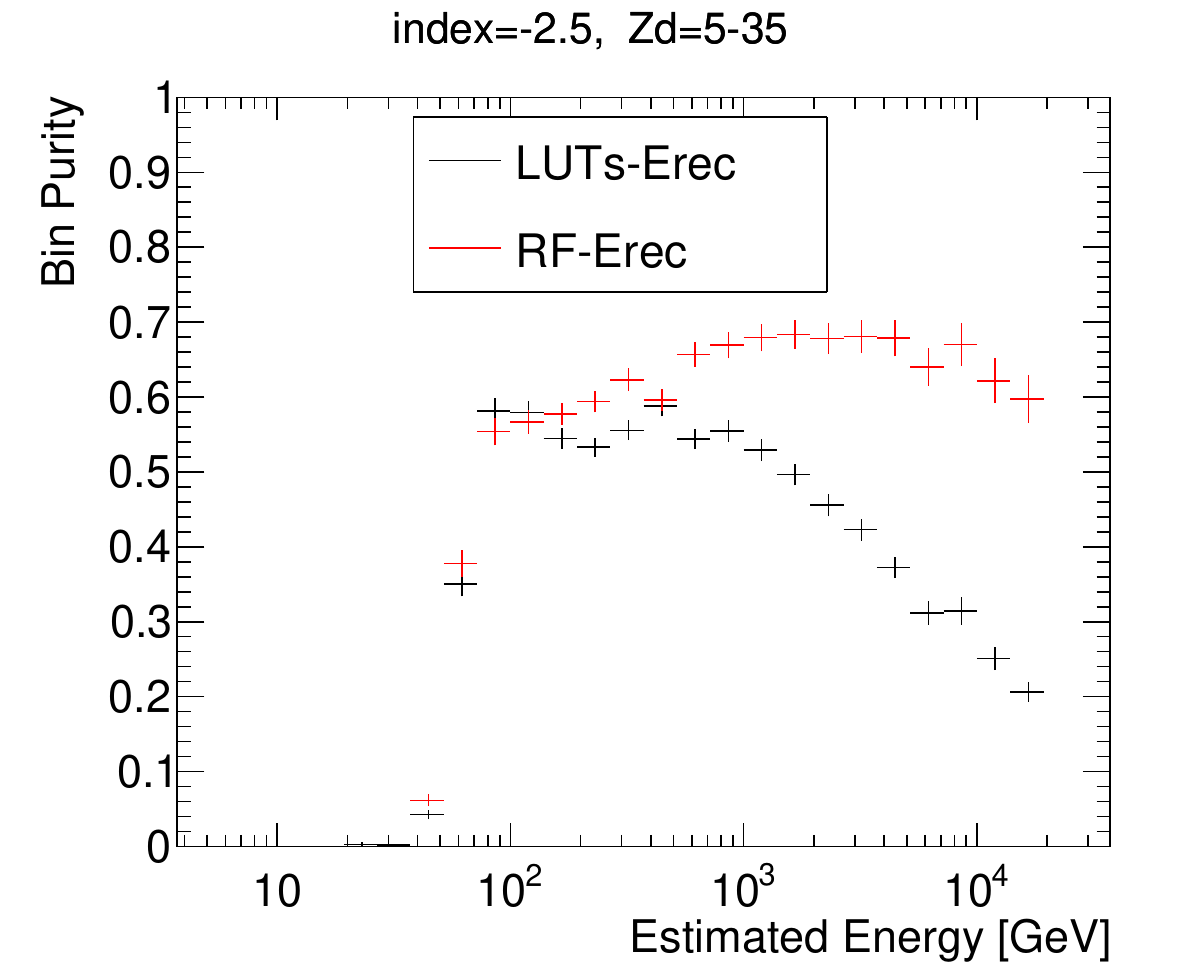} %{ReportBinPurity_LZd_1} 
\includegraphics[width=0.5\linewidth]%,pagebox=cropbox,clip,trim=0.5cm 0 0 0]
{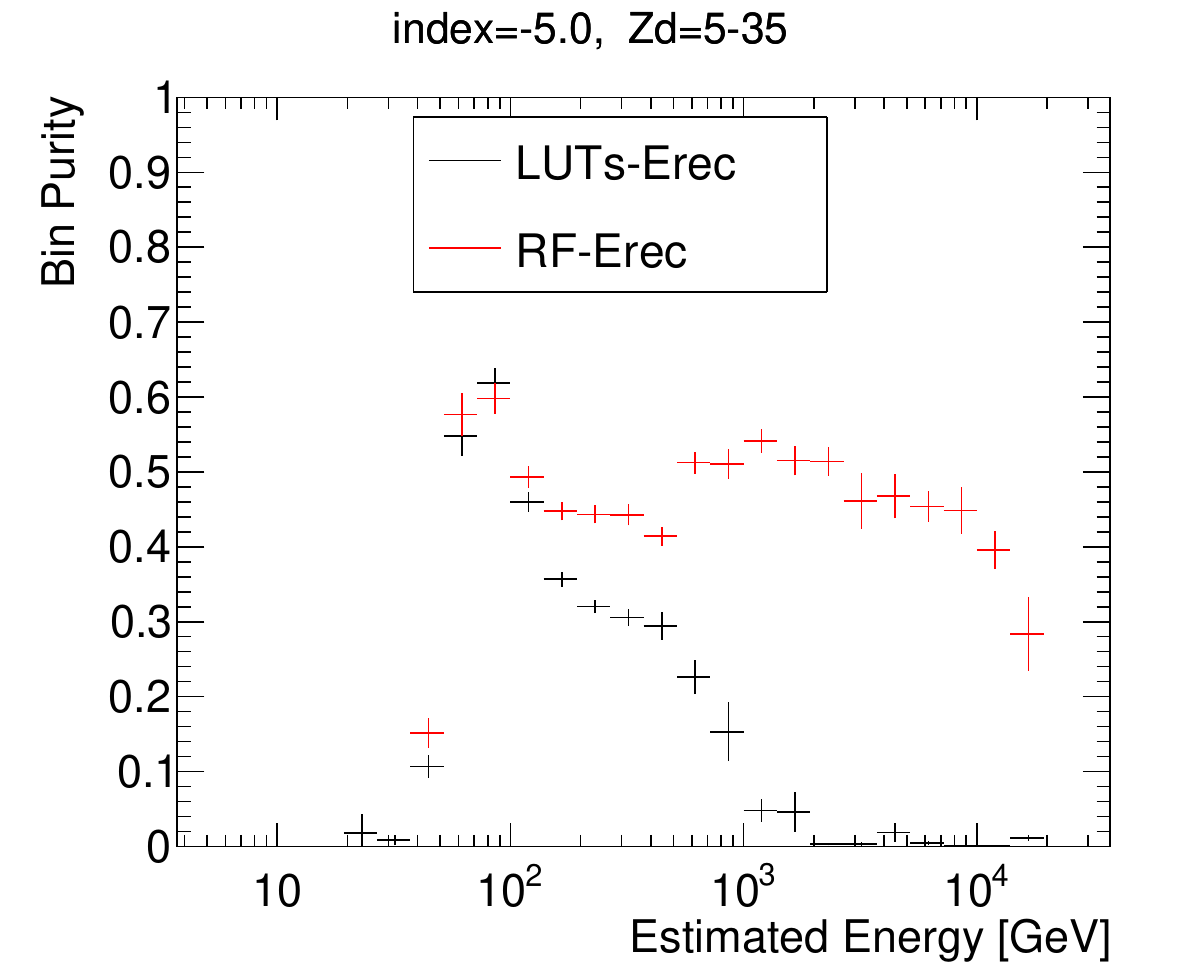}% {ReportBinPurity_LZd_2} 

\includegraphics[width=0.5\linewidth]%,pagebox=cropbox,clip,trim=0.5cm 0 0 0]
{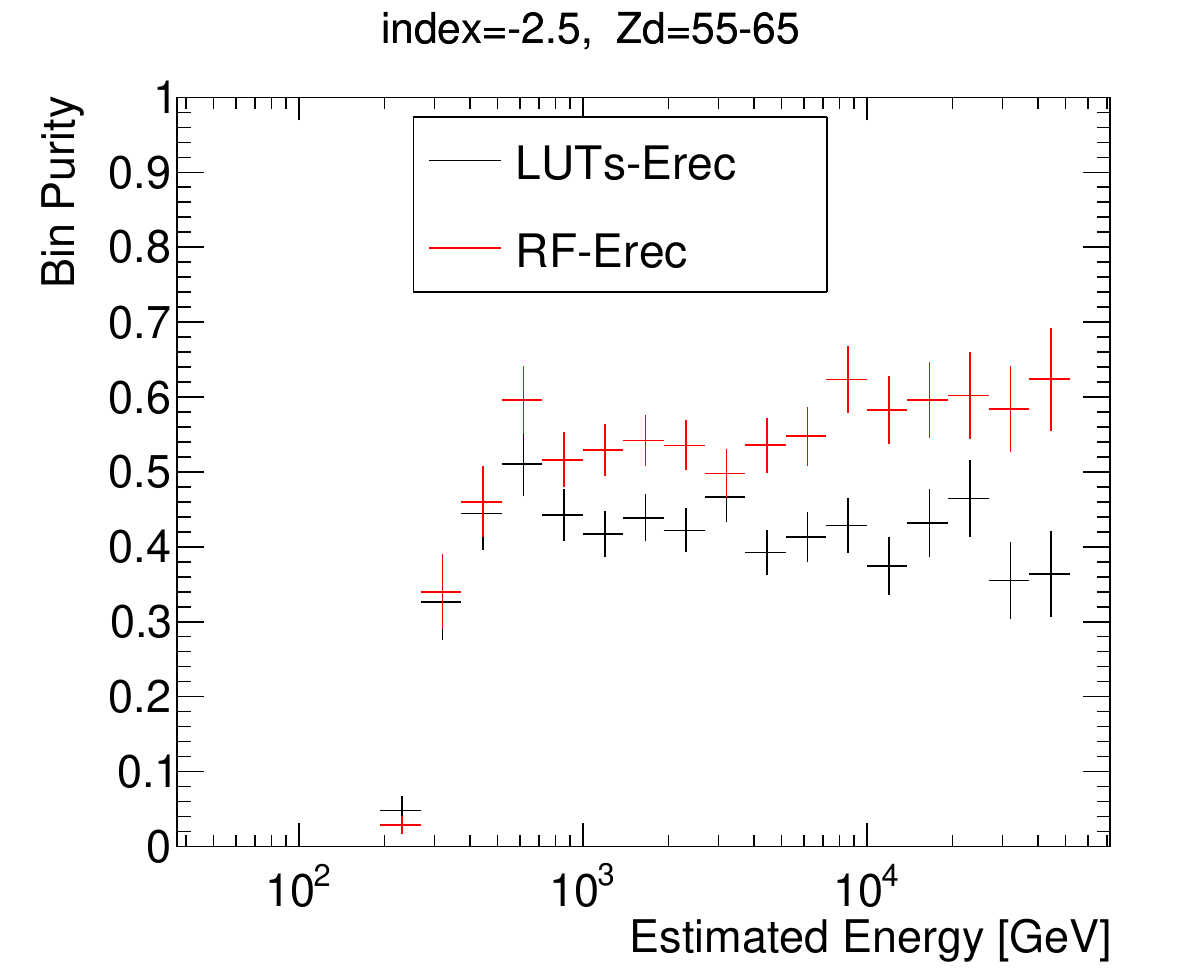}% {ReportBinPurity_HZd_1} 
\includegraphics[width=0.5\linewidth]%,pagebox=cropbox,clip,trim=0.5cm 0 0 0]
{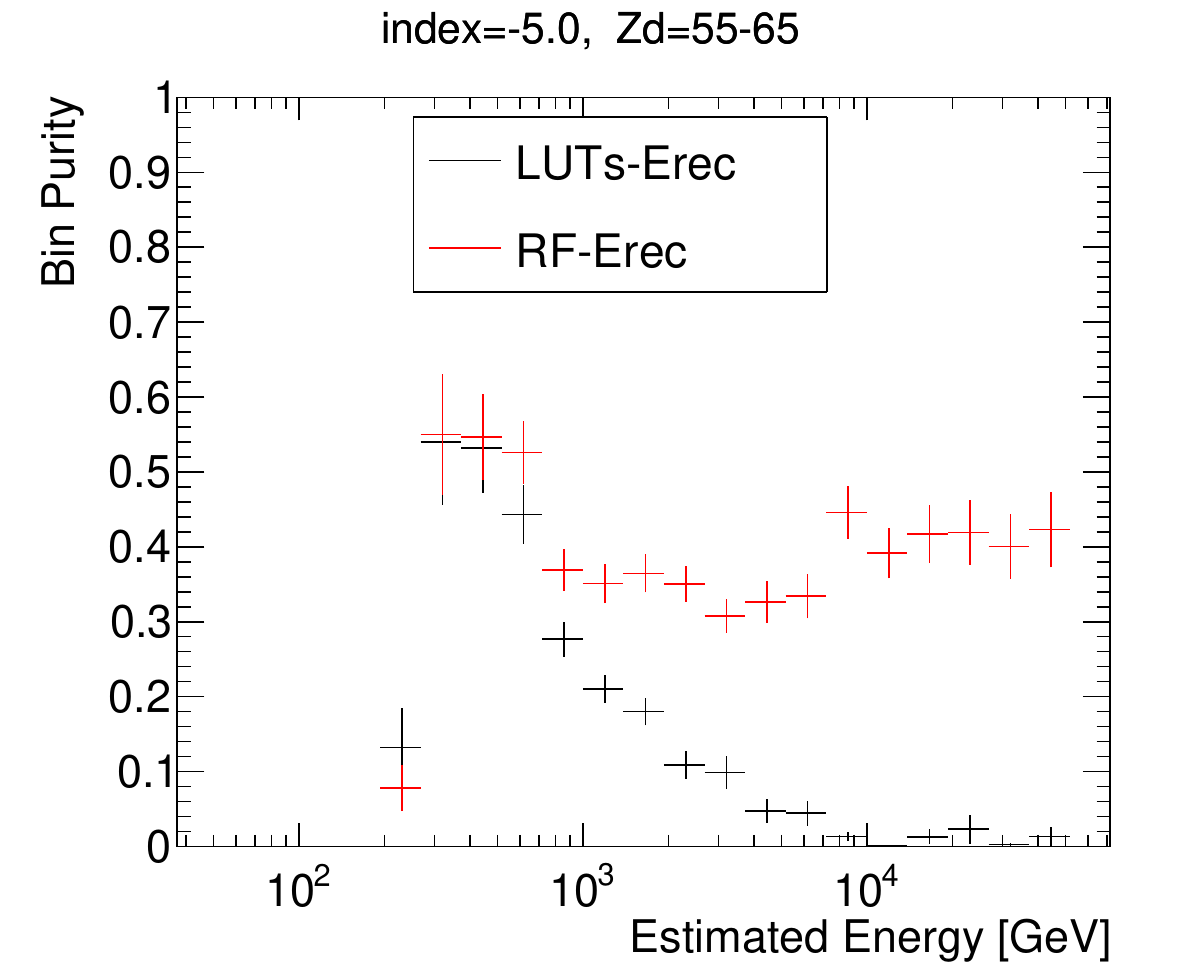}% {ReportBinPurity_HZd_2} 

\caption{
%Bin purity vs Energy with \mbox{LUTs-Erec} and \mbox{RF-Erec}
The bin purities as a function of energy, with \mbox{LUTs-Erec} and \mbox{RF-Erec}, are shown for two power-law indices and two Zd ranges, denoted in the panels. 
}\label{fig:Binpurity} 
\end{figure}

For any given estimated energy bin, we define bin purity as the fraction of events whose estimated energies remained in the same energy bin as their true energies. 
The result of this study is reported in \figref{fig:Binpurity}, and shows substantial differences between \mbox{\mbox{RF-Erec}} and \mbox{\mbox{LUTs-Erec}}, especially in the range above a few hundreds of GeV. The improved performance of \mbox{RF-Erec} is more dramatic in a steeper spectrum or a higher Zd observation, because the migration of events is largest.  
This comparison indicates that \mbox{RF-Erec} requires smaller corrections (smaller effect of the unfolding), and that it allows the study of gamma-ray source spectra with finer energy binning.

%------------------------------------------------
%  -- Generated by --
% /Users/kazuma/Workspace/MAGICana/macro_astro/EnergyMigrationEvaluator_Analytical
%------------------------------------------------

\subsubsection{Spillover}\label{sec:Spillover}
%%%%%%%%%%%%%%%
% 2022/06/06 at astro pc, plot_HardCutoffComparison_analytical.C
%%%%%%%%%%%%%%%
% The number of bins are aligned to be 7 per decade
%  (by using the same flute files)
% Cleaning migmat was once discussed but cancelled. 
%  (sys. eff. disappeared once the strategy shifted to analytical)
% 

The spillover is the migration of events from the correct energy bin (the bin that contains the true energy of the event) to a different (wrong) energy bin.  
In order to derive reliable gamma-ray spectra, this migration of events  needs  to be corrected with an unfolding procedure. And naturally, the bigger the migration of events to high energies, the more difficult is the correction of such an effect, and the higher the probability of having artificial (wrong) features in the final gamma-ray spectra. These spectral artifacts, such as multi-TeV gamma rays for a very distant source where no TeV photons are expected, may lead the analyzer to consider exotic physics explanations (e.g., presence of Axion Like Particles, or Lorentz Invariance Violation effects), and hence it is serious problem that needs to be treated carefully.

\begin{figure}[htbp]
\includegraphics[width=0.5\linewidth]%,pagebox=cropbox,clip,trim=0 0 0 0]
{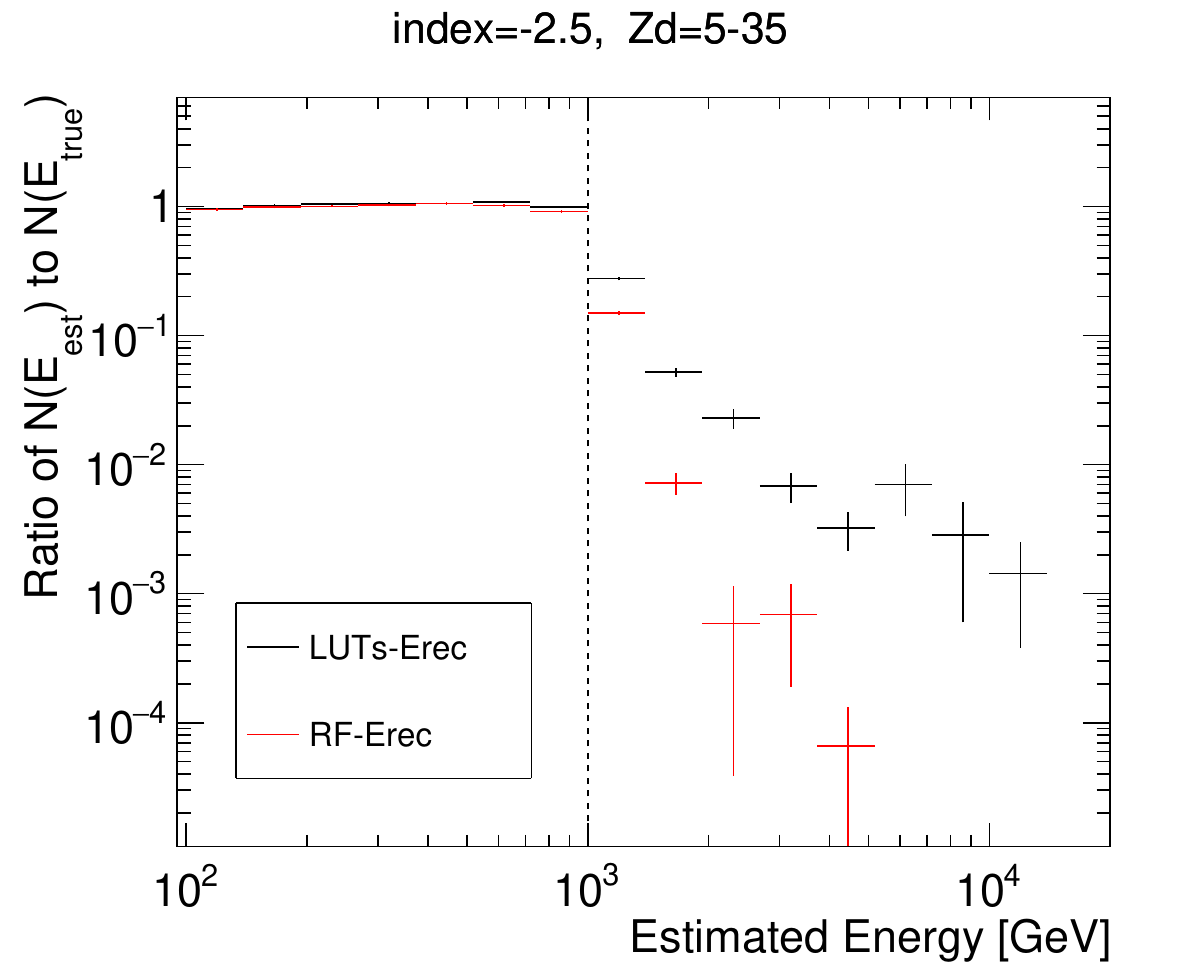}
\includegraphics[width=0.5\linewidth]%,pagebox=cropbox,clip,trim=0.5cm 0 0 0]
{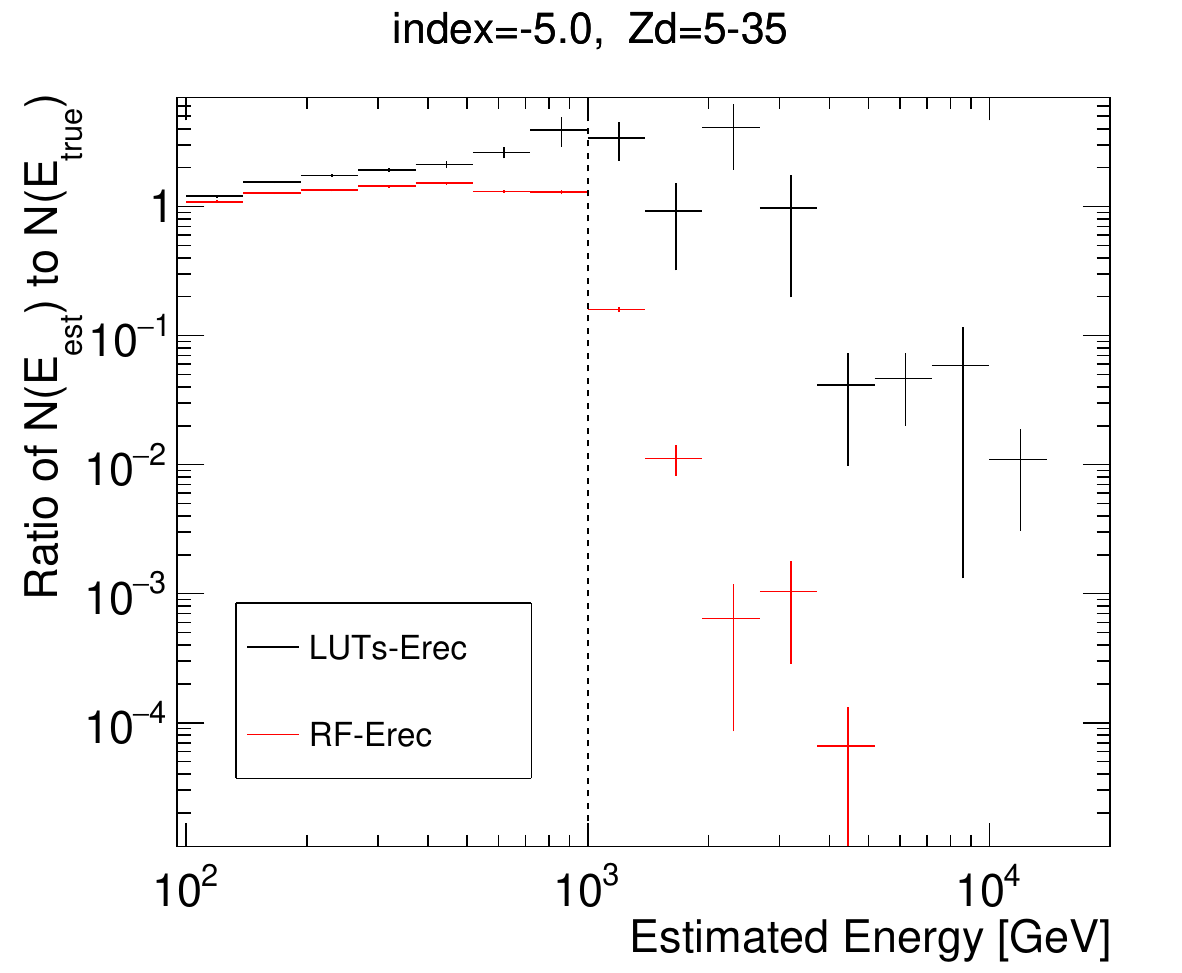}

\includegraphics[width=0.5\linewidth]%,pagebox=cropbox,clip,trim=0.5cm 0 0 0]
{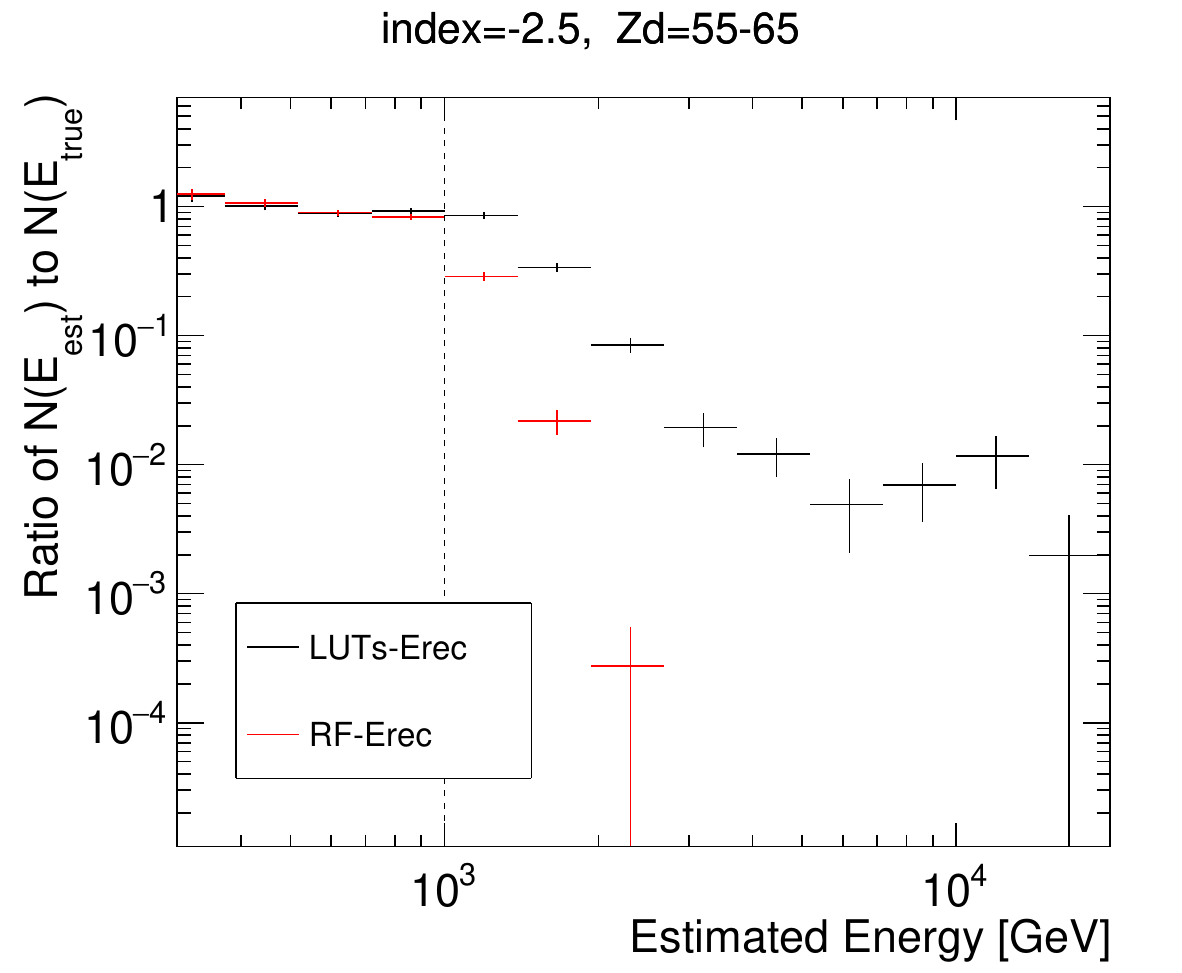}
\includegraphics[width=0.5\linewidth]%,pagebox=cropbox,clip,trim=0.5cm 0 0 0]
{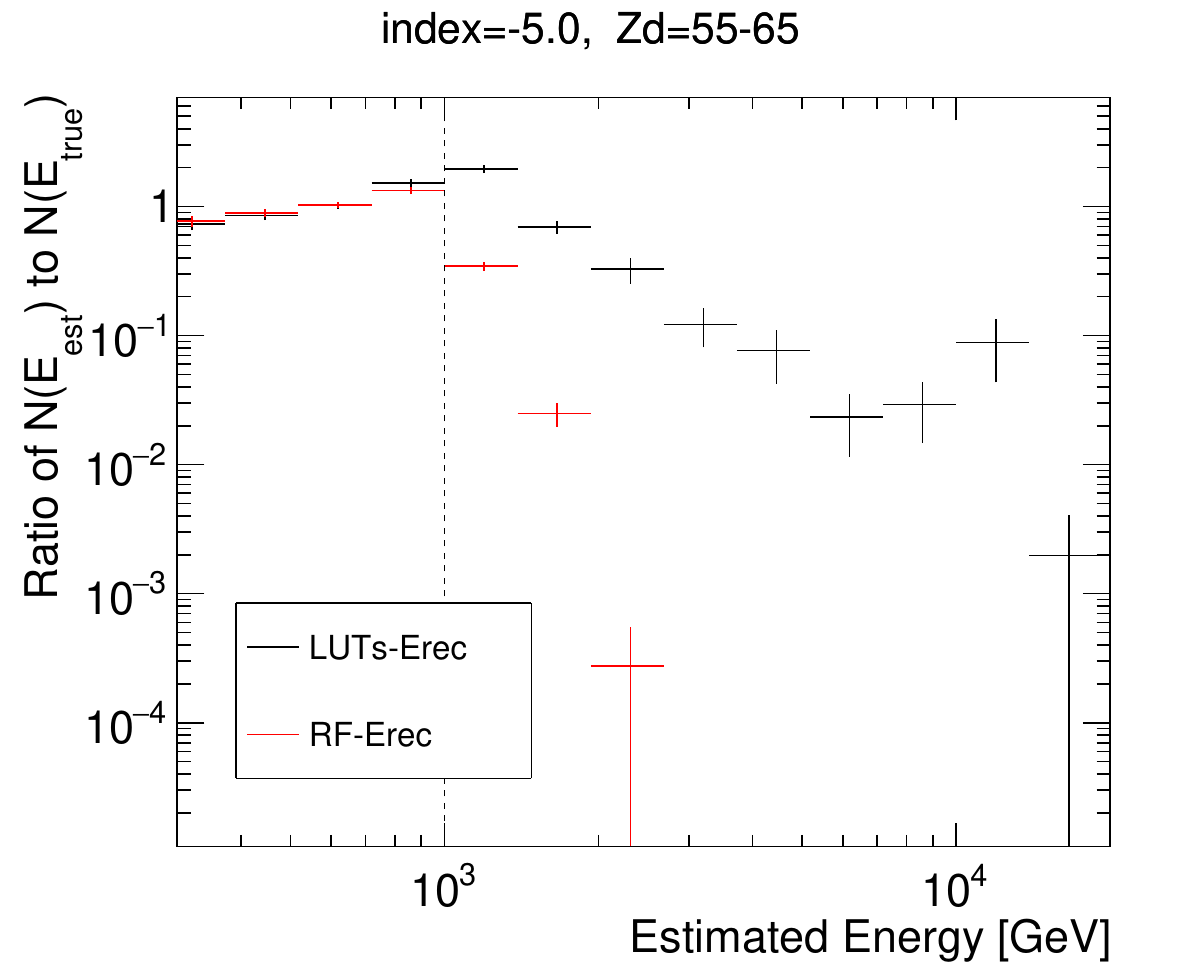}

% \includegraphics[width=0.95\textwidth,pagebox=cropbox,clip%,trim=0 0 0 2.1cm
% ]{HardCutoff_SpilloverComparison.pdf}%{HardCutoffComparison_20190530_3h_all}
\caption{ %Spillover ratio vs energy with \mbox{LUTs-Erec} and \mbox{RF-Erec}, for spectra with hard cutoff at 1 TeV
Spillover ratio vs energy with \mbox{LUTs-Erec} and \mbox{RF-Erec}, for power-law spectra with index -2.5 and -5.0, and with a hard cutoff at 1 TeV. The results are shown for a low Zd range (5-35deg) and a high Zd range (55-65deg).  The spillover ratio is calculated as the ratio of the number of events of $E_{est}$ in the same bin of $E_{true}$ to that of $E_{true}$. Owing to the hard cutoff, there is no event above \mbox{1 TeV} of true energy (pure spillover region), thus the spillover ratio is computed as the ratio to the number of events in the last bin below \mbox{1 TeV}. 
}\label{fig:HardCutoff} 
\end{figure}

In order to investigate these effects, we evaluated the spillover produced in power-law spectra with a hard cutoff at energy $E_{\rm cut}$= 1\,TeV. As we did for the evaluation of the bin purity, here we also used two distinct power-law index, typical (-2.5) and soft (-5.0), and considered two zenith angle ranges: low (5-35 deg) and high (55-65deg). 
After the generation of the respective simulated observations, the histograms of $E_{true}$  and $E_{est}$ are compared bin by bin, and the spillover ratio is computed, as the ratio of the number of events in $E_{est}$, denoted as $N(E_{est})$, to that in $E_{true}$, denoted as $N(E_{true})$. Above the cutoff energy of 1 TeV, there is no event in $E_{true}$, thus the denominator is set to the number of events (in $E_{true}$) at the last bin below 1 TeV. The results of this exercise are shown in \figref{fig:HardCutoff}.

%------------------------------------------------
%  -- Generated by --
% /Users/kazuma/Workspace/MAGICana/macro_astro/EnergyMigrationEvaluator_Analytical
%------------------------------------------------

%%% ---- Below 1 TeV ----- %%%

Below  and around 1 TeV, for the typical power-law spectrum (index=-2.5), the ratio is close to 1 for both \mbox{RF-Erec} and \mbox{LUTs-Erec}. On the other hand, for the soft spectrum (index=-5.0), the spillover increases and the ratio becomes larger than 1 for both energy reconstruction methods. In the case of \mbox{RF-Erec} the increase is small and always relatively close to 1, but for \mbox{LUTs-Erec}, the ratio increases beyond 2 around 1\,TeV, and indicates a much larger contamination from the "wrong" energies. This result is consistent with the drop of bin purity shown in \figref{fig:Binpurity}.

%  ----- Above 1 TeV  ----- 

As for the energy range above the cutoff energy of 1 TeV, an optimal energy reconstruction method should yield spillover ratios that are 0 because there is no real signal above this cutoff energy. The panels in \figref{fig:HardCutoff} show that, for all the conditions investigated,  \mbox{RF-Erec} yields a spillover ratio that quickly falls down above 1\,TeV, and essentially becomes 0 beyond 2--3\,TeV.  
 On the other hand, \mbox{LUTs-Erec} shows a long tail that extends even beyond 10\,TeV, with several \% of the events with true energies at the last bin below 1 TeV. Even though the number of events is small, these few events at 10 TeV and beyond would show up in the reconstructed spectra.  Owing to limited statistics of events from the observations, these features would be difficult to correct, and would stand out in the final measurements, forcing the scientists to consider exotic physiscs explanations.

\section{Summary and Concluding Remarks}
\label{sec:Conclusions}

%%% David Paneque (2023/06/24) %%%
%%% For better clarity, I decided to write them from scratch %%%%%

This manuscript reports the development of a novel methodology of energy reconstruction for very high energy gamma rays detected with Imaging Atmospheric Cherenkov Telescopes (IACTs). This methodology, based on the machine learning algorithm Random Forest, and named \mbox{RF-Erec}, has been adjusted for being used with data from the Major Atmospheric Gamma-ray Imaging Cherenkov (MAGIC) stereo telescope system, which is a worldwide leading instrument for gamma-ray astronomy in the energy range from about 20\,GeV to beyond 100\,TeV. However, in principle, the 
\mbox{RF-Erec} method could also be adjusted to work with data from any other IACTs.

Since MAGIC started to operate in stereo mode (fall of the year 2009), the energy estimation method that has been used is based on Look Up Tables that map the energy of the gamma ray and the values from the variables derived from the images in the camera of the telescopes. This method, which we name \mbox{LUTs-Erec}, has been the standard energy reconstruction method in the MAGIC collaboration, and used in over a 100 scientific publications by the MAGIC collaboration.

The superior capability of the newly developed method, RF-Erec, compared to LUTs-Erec, is visible in the shape of the migration matrix (see \figref{fig:MigMatrix}), which is substantially narrower and with smaller tails: the better-behaved migration matrix implies a smaller deformation caused by energy misreconstructions, and hence it eases the corrections that need to be applied to the gamma-ray spectra (e.g., forward folding). Specifically, RF-Erec improves the energy reconstruction performance in the following relevant aspects:

\begin{itemize}

\item  Improvement in energy resolution by a factor $\sim$2  \\
The energy resolution is typically quantiﬁed by the width of a Gaussian ﬁtted to the error distribution, which focuses on the majority (central part) of the distribution. We call it energy {\it resolution-$\sigma$}. The \mbox{RF-Erec} energy resolution-$\sigma$ is 20\% at 100 GeV, 11\% above 1 TeV in low Zenith distance (Zd), 20\% at 1\,TeV and 13\% above 10\,TeV in high Zd. For a wide range of the observable energies, the improvement of energy resolution-$\sigma$, compared to LUTs-Erec, reaches roughly factor of two, and the improvement is even larger for high Zd observations (see \figref{fig:GeneralizationErr}).

\item  Reduction in the energy errors dispersion by a factor $\sim$3 \\
In order to consider the tails  of the distribution of normalized errors, we also quantify the energy resolution with the standard deviation of the error distribution, which we name energy {\it resolution-SD}. For \mbox{RF-Erec}, the resolution-$\sigma$ and resolution-SD are very similar, indicating that the distribution of normalized errors is close to a Gaussian function, even far away from the central region. However, there is a large discrepancy between the resolution-$\sigma$ and the resolution-SD in the \mbox{LUTs-Erec}, which indicates large deviations from a Gaussian distribution, and hence the presence of large amount of outliers that worsen the measurement of the energy spectra of the gamma-ray sources.  The RF-Erec yield a resolution-SD that is about a factor of three better than that of LUTs-Erec (see \figref{fig:GeneralizationErr}).

\item Substantial reduction in the distortions of the gamma-ray spectra due to migration of events   \\
We quantify the migration of events with the {\it bin purity} and the {\it spillover}. The former is the fraction of gamma-ray events in a given estimated energy bin, whose true energies are also within the same energy bin. The latter is the migration of events from the correct energy bin (the bin that contains the true energy of the event) to a different (wrong) energy bin. The superior performance of RF-Erec shows up at a few hundred GeV, and specially above 1\,TeV. For a gamma-ray source with typical spectrum described with power-law index of -2.5, RF-Erec improves the bin purity above 1\,TeV by $\sim$50\%, while for a steep-spectrum described with a power-law index of -5.0, RF-Erec provides a bin purity that is more than one order of magnitude better than that of LUTs-Erec for energies above 1 TeV (see \figref{fig:Binpurity}). The spillover is quantified for a realistic case of a gamma-ray source with a power-law spectra that has a cutoff at 1\,TeV. We show that the spillover extends to at most factor of a few when using RF-Erec, while it extends to more than one order of magnitude when using LUTs-Erec (see \figref{fig:HardCutoff}). The superior performance of RF-Eref is particularly noticeable for steep power-law spectra (index -5.0) and high Zenith distance observations ($>$55deg). 

\end{itemize}

We want to notice that, in the literature, the energy resolution is often evaluated only in terms of the bias and the energy resolution, defined by the $\sigma$ value derived with a Gaussian fit or with the 68\% containment (i.e., resolution-$\sigma$). In the study we reported here, we also consider the energy dispersion (through resolution-SD) and the migration of events (through the bin purity in the spillover), which are very important for a reliable reconstruction of the energy spectra of the gamma-ray sources, specially those with steep power laws and/or with energy cutoffs. A large dispersion and migration of events may yield artificial spectral features that may lead to wrong scientific interpretations. These effects are particularly important at the high end of the gamma-ray spectra, where a few extra high-energy photons could have dramatic consequences for studies related to the EBL attenuation, Lorentz invariance violation, or searches for Axion-like-particles.

The RF-Erec has been proven to be reliable under a number of consistency tests, and carefully evaluated with data from the Crab Nebula, the standard candle for the VHE gamma-ray community. This new methodology has been validated by the MAGIC software board, implemented in the MAGIC Analysis and Reconstruction Software (MARS), and it is now the standard method adopted in analyses. It was already used in impactful scientific publications, like the characterization of the first GRB detected at TeV energies \citep{Acciari2019a, Acciari2019b}, and the related Lorentz Invariance Violation tests \citep{MAGICCollaboration2020}, where the superior performance of RF-Erec for steep-spectra was very relevant for the accurate and reliable determination of the results and their scientific interpretations.

\section{Acknowledgements}
We would like to thank the MAGIC collaboration for allowing us the usage of software tools, standard Monte Carlo data and observation data, that were needed to perform this study. We would especially like to thank Masahiro Teshima and Thomas Kuhr for supervising the work externally, and Julian Sitarek for numerous and fruitful discussions about the technicalities of the studies performed, as well as for a careful reading of the manuscript. Additionally, we also want to thank Koji Noda, Abelardo Molarejo and Darko Zaric for discussions that were very important for the initial investigations, and Chai Yating for many discussions and extensive studies towards better understanding of systematic uncertainties. The authors also want to thank the journal referees for very valuable and constructive remarks that helped improving the quality of the content and the text of this manuscript. K.I. acknowledges support from the Polish Ministry Of Education and Science grant No.2021/WK/08 and the Polish National Science Center grant No.2021/43/D/ST9/01153, and D.P. acknowledges support from the Deutsche Forschungsgemeinschaft (DFG; German Research Foundation) under Germany's Excellence Strategy EXC-2094—390783311.

\bibliographystyle{EnergyReconstructionMAGIC} 
\bibliography{EnergyReconstructionMAGIC.bib}


%\begin{thebibliography}{00}



%\end{thebibliography}
\end{document}